\documentclass[12pt, a4paper]{article}
\usepackage{graphicx}
\usepackage{amssymb}
\usepackage{amsmath}
\usepackage{bm}
\usepackage{xcolor}
\usepackage{theorem}
\usepackage{subcaption}
\usepackage{listings}
\usepackage{colortbl}
\usepackage{tabularx}
\usepackage{longtable}
\usepackage[utf8]{inputenc}
\usepackage[T1]{fontenc}
\usepackage{lmodern}
\usepackage[top=30truemm,bottom=30truemm,left=25truemm,right=25truemm]{geometry}
\usepackage[sort&compress,numbers, merge]{natbib}
\usepackage{multirow}
\usepackage{here}
\usepackage{braket}
\usepackage{colortbl}

\definecolor{Orange}{cmyk}{0,0.61,0.87,0}
\definecolor{JungleGreen}{cmyk}{0.99,0,0.52,0}
\definecolor{OliveGreen}{cmyk}{0.64,0,0.95,0.40}
\definecolor{Brown}{cmyk}{0,0.81,1,0.60}
\definecolor{RoyalBlue}{cmyk}{0.71,0.53,0,0.12}
\definecolor{Gray}{cmyk}{0,0,0,0.40}
\definecolor{LightPink}{cmyk}{0.0,0.25,0,0}
\definecolor{LLightPink}{cmyk}{0.0,0.10,0,0}
\definecolor{LightBlue}{cmyk}{0.25,0,0,0}
\definecolor{LightGray}{cmyk}{0,0,0,0.2}

\newcommand{\del}{\partial}

\newcommand{\nn}{\nonumber}

\usepackage[colorlinks=true, linkcolor=OliveGreen, citecolor=RoyalBlue,urlcolor=RoyalBlue]{hyperref}

\renewcommand{\arraystretch}{1.3}
\renewcommand{\thefootnote}{\fnsymbol{footnote}}
\allowdisplaybreaks[1]

\begin{document}

\begin{titlepage}
\begin{flushright}
\end{flushright}

\begin{center}

\vspace{1.5cm}

\textbf{\LARGE  Vortex Creep Heating in Neutron Stars}
\vspace{0.8cm}

Motoko Fujiwara$^{a,b}$\footnote{
 E-mail address: \href{mailto:motoko.fujiwara@tum.de}{\tt motoko.fujiwara@tum.de}},
Koichi~Hamaguchi$^{a,c}$\footnote{
 E-mail address: \href{mailto:hama@hep-th.phys.s.u-tokyo.ac.jp}{\tt hama@hep-th.phys.s.u-tokyo.ac.jp}}, 
Natsumi Nagata$^a$\footnote{
E-mail address: \href{mailto:natsumi@hep-th.phys.s.u-tokyo.ac.jp}{\tt natsumi@hep-th.phys.s.u-tokyo.ac.jp}},
\\
and 
Maura E. Ramirez-Quezada$^{a,d}$\footnote{
E-mail address: \href{mailto:me.quezada@hep-th.phys.s.u-tokyo.ac.jp}{\tt me.quezada@hep-th.phys.s.u-tokyo.ac.jp}},

\vskip 0.8cm

{\it $^a$Department of Physics, University of Tokyo, Bunkyo-ku, Tokyo
 113--0033, Japan} \\[2pt]
  {\it$^b$ Physik-Department, Technische Universit\"at, M\"unchen, James-Franck-Stra\ss e, 85748 Garching, Germany}\\[2pt] 
{\it $^c$Kavli IPMU (WPI), University of Tokyo, Kashiwa, Chiba
 277--8583, Japan} \\[2pt]
 {\it $^d$Dual CP Institute of High Energy Physics, C.P. 28045, Colima, M\'exico}
 
\date{\today}

\vskip 1.5cm

\begin{abstract}
  Recent observations of old warm neutron stars suggest the presence of a heating source in these stars, requiring a paradigm beyond the standard neutron-star cooling theory. In this work, we study the scenario where this heating is caused by the friction associated with the creep motion of neutron superfluid vortex lines in the crust. As it turns out, the heating luminosity in this scenario is proportional to the time derivative of the angular velocity of the pulsar rotation,
  and the proportional constant $J$ has an approximately universal value for all neutron stars. This $J$ parameter can be determined from the temperature observation of old neutron stars because the heating luminosity is balanced with the photon emission at late times. 
  We study the latest data of neutron star temperature observation and find that these data indeed give similar values of $J$, in favor of the assumption that the frictional motion of vortex lines heats these neutron stars. These values turn out to be consistent with the theoretical calculations of the vortex-nuclear interaction. 
\end{abstract}
\end{center}
\end{titlepage}

\renewcommand{\thefootnote}{\arabic{footnote}}
\setcounter{footnote}{0}

\section{Introduction}

A neutron star (NS) is incredibly dense and exists under extreme conditions of pressure and temperature that cannot be found in other places in the universe. While the internal structure of NSs remains elusive, indirect evidence suggests the existence of a neutron superfluid in the inner crust. The first indication of superfluidity in NSs came from the observation of non-zero pairing energy associated with attractive forces, leading to the formation of an energy gap and hence superfluidity~\cite{Bohr:1958zz}. This scenario was predicted to occur in stars with neutron cores~\cite{MIGDAL1959655}. After the discovery of pulsars, the energy gap in NS matter has been further studied; see, \textit{e.g.}, Refs.~\cite{Page:2013hxa, Haskell:2017lkl, Sedrakian:2018ydt} for a recent review on superfluidity in NSs. 

In a rotating NS, the irrotational property of a superfluid requires the formation of vortex lines, whose distribution determines the angular velocity of the superfluid component. In the inner crust region, these vortex lines are fixed at certain positions by the interactions with nuclei and cannot move freely. This preserves the rotational speed of the superfluid component and prevents it from following the slowdown of the pulsar rotation, giving rise to the deviation in the rotational speed between the superfluid and other components. This deviation increases until the vortex lines are forced to move by the Magnus force, which increases as the difference in the rotational speed increases. This vortex-line dynamics leads to some observational consequences. A well-known example is the pulsar glitch phenomenon, namely, sudden changes in the rotational frequency of NSs,\footnote{See Refs.~\cite{Haskell:2015jra, Fuentes:2017bjx, Antonopoulou:2022rpq, Zhou:2022cyp} for a recent review on pulsar glitches. } which could be attributed to an avalanche of unpinning of superfluid vortex lines~\cite{Anderson:1975zze, 1977ApJ...213..527A}. Another phenomenon is the heating effect caused by the friction associated with the creep motion of vortex lines~\cite{1984ApJ...276..325A, 1984ApJ...278..791A, 1985ApJ...288..191A, 1989ApJ...346..808S, 1989ApJ...346..823A, 1991ApJ...381L..47V, 1993ApJ...408..186U, 1995ApJ...448..294V, Schaab:1999as, Larson:1998it, Gonzalez:2010ta}, which is the subject of the present paper. 

Our motivation to revisit this heating mechanism is provided by the recent observations of old warm NSs~\cite{Kargaltsev:2003eb, Mignani:2008jr, Durant:2011je, Rangelov:2016syg, Pavlov:2017eeu, Abramkin:2021tha, Abramkin:2021fzy}, whose observed temperature is considerably higher than that predicted in the standard NS cooling scenario~\cite{Yakovlev:1999sk, Yakovlev:2000jp, Yakovlev:2004iq, Page:2004fy, Gusakov:2004se, Page:2009fu}. This work aims to study whether these observations can be explained by the vortex-creep heating effect. With this objective in mind, we focus on the following characteristic property of the vortex-creep heating. As we see below in detail, the heating luminosity in this heating mechanism is proportional to the time derivative of the angular velocity of the pulsar rotation, and the proportional constant is determined only by the NS structure and the vortex-nuclear interactions. As a result, the value of this proportional constant, denoted by $J$ in this paper, is almost universal over NSs. In addition, we can obtain the value of $J$ for an old NS by observing its temperature and its pulsar motion since, at late times, the heating luminosity balances with the luminosity of photon emission, which is determined by the surface temperature of the NS. As it turns out, the present data of old warm NSs indeed show similar values of $J$ for these stars, in agreement with the prediction of the vortex-creep heating scenario. We also find that these values are compatible with the $J$ parameter evaluated from the calculations of the nuclear pinning force available in the literature.

The remainder of this paper is structured as follows. In Section~\ref{sec:thermal_evolution}, we provide a brief overview of the thermal evolution of NSs, focusing on the isothermal phase. In Section~\ref{sec:vortex_creep}, we describe the vortex creep heating mechanism for NSs and explain how we can relate a universal parameter with the late-time temperature prediction.
In Section~\ref{sec:prediction}, we summarize the numerical evaluation of the pinning force, from which we calculate the parameter $J$. In Section~\ref{sec:observation}, we study the recent observations of old and warm NSs to assess the current status of the vortex creep heating hypothesis. Finally, we summarize our findings and conclude our discussion in Section~\ref{sec:conclusion}.

\section{Thermal evolution of neutron star}
\label{sec:thermal_evolution}

This section reviews the surface temperature prediction of NSs based on their thermal evolution and sources of heating and cooling. The temperature distribution within NSs is characterized by a local core temperature $T$, which exhibits a temperature gradient only during the early stages of the NS, typically within the first 10--100 years of its existence~\cite{1994ApJ...425..802L, Gnedin:2000me}. The high thermal conductivity of the highly degenerate electron gas causes the core to become isothermal over time, reaching thermal equilibrium. Hence, the red-shifted temperature of NSs, $T^\infty(\bar{r},t)=T(\bar{r},t) e^{\phi(\bar{r})}$, where $\phi(\bar{r})$ specifies the gravitational redshift, reaches a constant value $T^\infty(\bar{r},t)\simeq T^\infty(t)$ and only the outermost layers exhibit an appreciable temperature gradient. The relativistic equation describing the thermal evolution of NSs after thermal relaxation is given by~\cite{1977ApJ...212..825T,1980ApJ...239..671G}
\begin{eqnarray}
  C(T^\infty)\frac{dT^\infty}{dt}=-L_\nu^\infty(T^\infty)-L_\gamma^\infty(T^\infty)+L_{\rm H}^\infty.\label{eq:Tevol}
\end{eqnarray}
Here, $C$ represents the total heat capacity of the NS and is temperature-dependent. The right-hand side of the equation expresses the red-shifted luminosity for three different processes, namely, neutrino cooling $L_\nu^\infty$, photon cooling $L_\gamma^\infty$, and heating source $L_{\rm H}^\infty$.

At temperatures below a few $\times~10^9~\mathrm{K}$, the NS becomes transparent to neutrinos, allowing them to escape without interacting with the stellar matter and carrying away energy. Therefore, the cooling process during the early stage of the star's life is dominated by neutrino emission. At later times, typically for $t \gtrsim 10^5~\mathrm{yrs}$,\footnote{The dominance of photon emission might be delayed for massive NSs, in which the rapid neutrino emission via the direct Urca process could occur; we do not consider this possibility in what follows. } photon emission dominates neutrino emission. The thermal photon emission follows a blackbody spectrum. The photon luminosity $L_\gamma$ can be described by the Stefan-Boltzmann law and related to surface temperature $T_{\rm  s}$ as
\begin{align}
    L_\gamma = 4\pi R_{\rm NS}^2\sigma_{\rm  SB}T_{\rm  s}^4,
    \label{eq:Lgamma_eq_LH}
\end{align}
in the local reference frame of the NS, where $\sigma_{\rm  SB}$ is the Stephan-Boltzmann constant and $R_{\rm NS}$ the NS radius. To relate the internal temperature $T$ of a NS to its surface temperature $T_{\rm  s}$, we use the heat envelope model proposed by Potekhin \textit{et al.} in 1997~\cite{Potekhin:1997mn}. According to this model, the observed thermal emission from old isolated NSs can be explained by the heat trapped in a thin envelope surrounding the star's crust. 

If we have an internal heating source, the heating luminosity will balance with the photon luminosity at a sufficiently late time,  
\begin{align}
  L^\infty_{\rm  H}  &\simeq  L^\infty_\gamma,
  \label{eq:L_g=L_H}
\end{align}
which determines the surface temperature. It is pointed out that NSs have some internal heating mechanisms, such as vortex creep heating, rotochemical heating, and magnetic field decay. These heating effects may become visible at late times and can operate even for isolated NSs. These internal mechanisms are comprehensively compared with the observed surface temperature in Refs.~\cite{Schaab:1999as, Gonzalez:2010ta}, which is recently revisited by Ref.~\cite{Kopp:2022lru}.

We note in passing that the balance equation~\eqref{eq:L_g=L_H} generically holds with good accuracy for old NSs that are older than $10^5~\mathrm{yrs}$. This can be seen by estimating the typical timescale $\tau_{\rm  eq}$ for the relaxation into the equilibrium state:
\begin{align}
  \tau_{\rm  eq}  &\simeq  
  \frac{C T^\infty}{4  \pi  R_{\rm  NS}^2  \sigma_{\rm  SB}  
  T_{\rm  s}^4}
  \\
  &\sim  
  3 \times  10^4~\mathrm{yrs}
  \left( \frac{C}{10^{35}~\mathrm{erg/K}} \right)
  \left( \frac{R_{\rm  NS}}{11.43~{\rm  km}} \right)^{-2}
  \left( \frac{T_{\rm  s}}{10^5~\mathrm{K}} \right)^{-4}
  \left( \frac{T^\infty}{10^6~\mathrm{K}} \right)
  ~,
\end{align}
where we have used $T^\infty\sim 10^6~\mathrm{K}$ corresponding to the surface temperature, at the equilibrium phase,  of  $T_{\rm  s}\sim 10^5~\mathrm{K}$.\footnote{To derive the typical scale of $\tau_{\rm  eq}$, we fitted the relation between the NS internal temperature $T$ and the NS surface temperature $T_{\rm  s}$ for $t_{\rm  age}  \lesssim  10^6~\mathrm{yrs}$~\cite{Potekhin:1997mn} by assuming $T  \propto T_{\rm  s}^2$.} It is found that this timescale is shorter than the age of old NSs, assuring the equilibrium condition~\eqref{eq:L_g=L_H}.

\section{Review of vortex creep heating}
\label{sec:vortex_creep}

In this section, we will review vortex creep heating, where the presence of a superfluid in the inner crust of a NS plays a key role. In this region, vortex lines are thought to exist as a consequence of the NS rotation. In Sec.~\ref{sec:EoM}, we will derive the equation of motion for this rotational motion by introducing two different angular velocities for the inner crust superfluid and the other part of the star. In Sec.~\ref{sec:dynamics}, we will consider the dynamics of a vortex line and evaluate its radial velocity. Finally, in Sec.~\ref{sec:surface_temperature}, we will assess the effect of vortex creep on the late-time temperature prediction of NSs.

\subsection{Equations of motion for crust and superfluid}
\label{sec:EoM}

To describe a rotating NS, let us divide it into two components depending on how external torque exerts on it, the \textit{crust component} and the \textit{superfluid component}~\cite{1969Natur.224..872B}. The crust component comprises a rigid crust, a lattice of nuclei, and charged particles tightly coupled to electromagnetic field lines~\cite{1969Natur.224..872B}. This component is directly affected by external torque provided by pulsar magnetic radiation. The superfluid component refers to $^1S_0$ superfluid of neutrons.\footnote{The neutron triplet ($^3P_2$) superfluid in the core region is classified into the crust component in the two-component model~\cite{1969Natur.224..872B}. This is because the neutron superfluid in this region is expected to coexist with the proton superconductor and be tightly coupled to the crust component~\cite{PhysRevD.25.967,1984ApJ...282..533A}.}
This superfluid phase is believed to appear in the inner crust region based on pairing gap evaluations~\cite{Page:2013hxa, Haskell:2017lkl, Sedrakian:2018ydt}. 
This component is just indirectly affected by the external torque through the interaction with the crust component. 
\begin{figure}[tb]
\begin{center}
\centering
\includegraphics[width=1\textwidth]{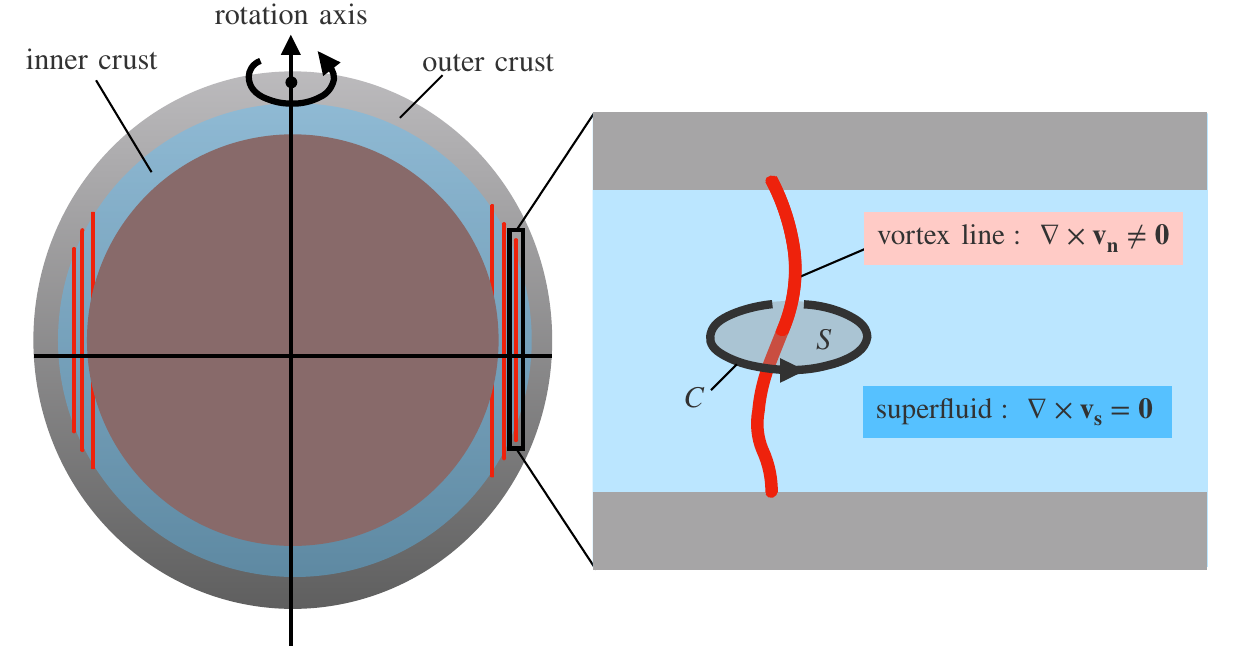}
\end{center}
\caption{The structure of a NS and vortex lines.
\textit{Left}: The grey, blue, and brown regions represent the outer crust, inner crust, and core regions, respectively. 
The red lines represent vortex lines. 
\textit{Right}: A single vortex line in the inner crust. The vortex line in the neutron superfluid is attached to the outer crust and has two boundaries.
}
\label{fig:vortex_line}
\end{figure}
On the left side of Fig.~\ref{fig:vortex_line}, we show a schematic diagram of a NS. The grey region is the outer crust, composed of ions and electrons. The blue layer represents the inner crust region and has an approximately $\sim  1~\mathrm{km}$ thickness. In this region, nuclei, electrons, and neutrons exist, and neutrons are expected to be in the form of the neutron singlet superfluid. The brown region is the NS core, whose internal structure remains uncertain and a subject of ongoing research and debate. 
In the two-component model, the neutron superfluid in the blue layer is classified into the superfluid component, while all the other parts are classified into the crust component. 

Based on this two-component model, we derive the equations of motion and describe the rotation. For later convenience, we divide the system into a thin disc and introduce the cylindrical coordinate $(r, \varphi, z)$. Since we treat the crust component as a rigid body, its angular velocity $\Omega_{\rm c}(t)$ is independent of $r$. On the other hand, the superfluid angular velocity varies with $r$, and we denote it by $\Omega_{\rm s}(t,r)$. The equation of motion for the crust component is 
\begin{align}
  I_{\rm  c}  \dot{\Omega}_{\rm  c} (t)  =  N_{\rm  ext} (t) +  N_{\rm  int} (t),
  \label{eq:EoM_crust}
\end{align}
where  $I_{\rm c}$ represents the moment of inertia of the crust component, and the dot represents the derivative with respect to time $t$. On the right-hand side, we divide the torque into two parts: The first term $N_{\rm ext} (t)$ represents the external torque acting on the system. The second term $N_{\rm  int} (t)$ corresponds to the effects of internal torques in the system
\begin{align}
N_{\rm int}(t) &= -\int d I_p(r) \frac{\partial \Omega_{\rm s}}{\partial t}(t,r),
\label{eq:nint}
\end{align}
where $dI_p$ represents the differential inertial momenta. The integral is taken over the region where the crust component interacts with the superfluid component, called the \textit{pinning} region. These two components connect through the pinning force between the vortex line and the nuclei-like object in the inner crust, as we will discuss in Sec.~\ref{sec:dynamics}. 

 The equations of the superfluid component are obtained by introducing two fundamental measures of rotation in the fluid: {vorticity} and {circulation}. The vorticity vector $\bm{\omega}$ characterizes the microscopic features of rotation and is a locally determined value defined as the curl of the fluid velocity $\bm{v}$,
\begin{align}
\bm{\omega} \equiv \nabla \times \bm{v}.
\end{align}
In contrast, the circulation $\Gamma$ measures the macroscopic rotation and is defined over a finite region. It is given by the line integral of the fluid velocity $\bm{v}$ around a closed path $C$, which can be expressed as a surface integral over any surface $S$ 
with the boundary $C$,
\begin{align}
\Gamma \equiv \oint_C \bm{v} \cdot d \bm{\ell} = \iint_S \bm{\omega} \cdot d \bm{S},
\label{eq:circulation}
\end{align}
where $d\bm{\ell}$ and $d\bm{S}$ denote the line and surface elements, respectively. We used Stokes' theorem to obtain the last expression.

Superfluid motion obeys the potential flow condition, 
\begin{align}
\nabla \times \bm{v}{_{\rm s}} = \bm{0},
\label{eq:potential_condition}
\end{align}
where $\bm{v}{_{\rm s}}$ denotes the superfluid velocity. This condition, implying the absence of the vorticity in the superfluid, holds because the superfluid velocity is proportional to the gradient of the phase of the condensate wave-function of the superfluid. Nevertheless, we still have a nonzero circulation if there exists a singular object known as a vortex line. In Fig.~\ref{fig:vortex_line}, we show a schematic picture of a single vortex line in the NS inner crust. The vortex line is a string-like configuration with a thickness of the order of femtometers (red curve).
The circulation for each vortex line is quantized in units of  
\begin{align}
  \kappa  &\equiv  \frac{h}{2  m_n},
\end{align}
where $h$ is the Planck constant and $m_n$ is the neutron mass. This quantization follows from the condition that the wave-function of the condensate is single-valued, and thus the change in its phase must be $2  \pi  k$ with $k$ as an integer.
Since a vortex line with $k=1$ is energetically favored and stable, the system with larger angular velocity contains a larger number of vortex lines~\cite{Lifshitz-Pitaevskii}. The number of vortex lines will be saturated if the total circulation reaches that of the rigid rotation as a whole system, which is expected in NSs. The vortex lines in the inner crust have boundaries corresponding to the normal matter in the outer crust. Under this circumstance, the circulation for the contour $C$ in Fig.~\ref{fig:vortex_line} is uniquely determined and, because of the potential flow condition~\eqref{eq:potential_condition}, can be regarded as topological as it remains unchanged under deformations of $C$ (unless it passes through another vortex line). 

We may express the superfluid velocity on average in the same form as the normal fluid,\footnote{This relation is confirmed by observing the shape of the free surface for rotating superfluid in liquid He system~\cite{PhysRevLett.33.280,Osborne_1950,PhysRevLett.69.2392}.}
\begin{align}
  \Braket{\bm{v}_{\rm  s}}  =  \bm{\Omega}_{\rm  s}  \times  \bm{r}~,
  \label{eq:v_s-average}
\end{align}
where $\bm{r}$ denotes the position vector from the center of the NS. The total circulation of a superfluid system is equal to the sum of the circulation of each vortex line. This means that the number of vortex lines is directly related to the superfluid angular velocity $\bm{\Omega}_{\rm s}$. By substituting Eq.~\eqref{eq:v_s-average}  in  Eq.~\eqref{eq:circulation}, we obtain 
\begin{align}  
  \Gamma_{\rm  superfluid}
  &=  \int_C  d  \bm{\ell}  \cdot  \bigl( \bm{\Omega}_{\rm  s}  (t, r)  \times  \bm{r} \bigr)
  =  \int_0^r  d  r'  ~  2  \pi  r'  \kappa  n  ( t,  r' ), 
  \label{eq:circulation_compare}
\end{align}
where we choose the integral path $C$ around the edge of the disc with radius $r$, and $n ( t,  r )$ is the number density of the vortex lines per unit area. Noting that only the radial motion of vortex lines changes $n$ due to the axial symmetry around the rotation axis, we obtain the following conservation law, 
\begin{align}
  \frac{\del  n}{\del  t}  +  \nabla  \cdot  ( n  v_r  \bm{e}_r )  =  0,
  \label{eq:number_cons}
\end{align}
where $v_r$ is the vortex velocity in the radial direction $\bm{e}_r$, which we call the \textit{creep rate}.  
Combining Eqs.~\eqref{eq:circulation_compare}~and~\eqref{eq:number_cons}, we obtain the equation of motion for the superfluid component:
\begin{align}
  \frac{\del  \Omega_{\rm  s}}{\del  t}  
  &=  
  -  \left( 2  \Omega_{\rm  s}  +  r  \frac{\del  \Omega_{\rm  s}}{\del  r} \right)  \frac{v_r}{r}.
  \label{eq:EoM_SF}
\end{align}
The NS rotation is described by Eqs.~\eqref{eq:EoM_crust}~and~\eqref{eq:EoM_SF}
coupled through a nonzero $v_r$. In other words, by switching off the radial motion of the vortex line, we have $\del  \Omega_{\rm  s} / \del  t  =  0$. 
In this case, $N_{\rm  int}$ turns out to be zero, and the two equations of motion are decoupled.

\subsection{Dynamics of a vortex line}
\label{sec:dynamics}

In the inner crust, a vortex line feels two forces, the \textit{pinning force} and the \textit{Magnus force}. The pinning force arises from the interaction between a vortex line and a nuclear-like object within the inner crust, resulting in the pinning of the vortex line to the lattice of nuclei where the energy is minimized~\cite{Anderson:1975zze}. As long as the pinning force is dominant, the vortex lines remain attached to the crust and move at the same velocity as the crust component: 
\begin{align}
  \bm{v}_{\rm  VL} (t, r)  =  \bm{\Omega}_{\rm  c} (t)  \times  \bm{r},
  \label{eq:v_VL}
\end{align}
where $\bm{v}_{\rm  VL}  (t,r)$ denotes the velocity of a vortex line. 

One way to quantify the pinning force is to compare the energies associated with different configurations of a vortex line.
\begin{figure}[tb]
\begin{center}
\centering
\includegraphics[width=0.8\textwidth]{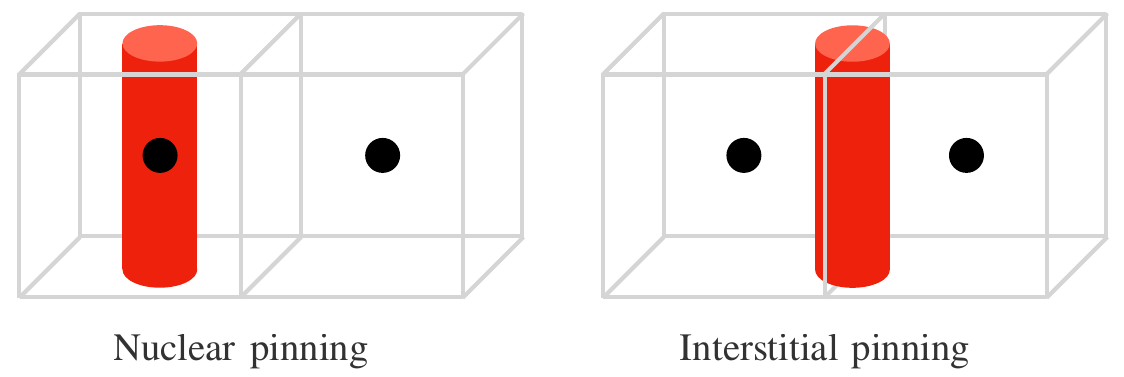}
\end{center}
\caption{
  The nuclear pinning and interstitial pinning for the vortex line pinning configurations. Red cylinders and black spheres represent vortex lines and nuclei, respectively. 
  }
\label{fig:configuration}
\end{figure}
In Fig.~\ref{fig:configuration}, we show two possibilities for the pinning configurations, 
the \textit{nuclear pinning} and  the \textit{interstitial pinning}. The difference in the energies between these two configurations defines the pinning energy, 
\begin{align}
  E_{\rm  pin}  \equiv  E_{\rm  NP}  -  E_{\rm  IP},
  \label{eq:Epin}
\end{align}
where $E_{\rm  NP}$ ($E_{\rm  IP}$) denotes the energy of the nuclear (interstitial) pinning configuration. 
A nuclear pinning configuration occurs when the pinning energy is negative, and the vortex line is directly attached to the nuclei lattice. Conversely, when the pinning force is repulsive, the vortex line is pinned in the interstitial regions. For a nuclear pinning configuration, a rather crude estimate of the pinning force per unit length is then given by 
\begin{align}
  f_{\rm  pin}|_{\rm  NP}  
  &\simeq 
  \frac{|E_{\rm  pin}|}{\Delta r  \Delta L }, 
  \label{eq:fpin_NP}
\end{align}
where $\Delta r$ is the distance between the nuclear and interstitial pinning positions and $\Delta L$ is the distance between the successive pinning sites along a vortex line. These two quantities are expected to be of the order of the Wigner-Seitz radius $R_{\rm  WS}$,  the radius of an imaginary sphere whose volume is equal to the average volume per nucleus in each region. 

Due to this pinning effect, the relative velocity between the superfluid and the vortex lines is developed,
\begin{align}
  \delta  \bm{v}  
  \equiv  \bm{v}_{\rm  s}  -  \bm{v}_{\rm  VL}
  =  \delta  \bm{\Omega}  \times  \bm{r}
  ,
  \label{eq:delta_v}
\end{align}
where we use Eqs.~\eqref{eq:v_s-average}~and~\eqref{eq:v_VL} to obtain the last expression and introduce the relative angular velocity, 
\begin{align}
  \delta  \bm{\Omega}  &\equiv  \bm{\Omega}_{\rm  s}  -  \bm{\Omega}_{\rm  c}.
\end{align}
This velocity difference induces the Magnus force per unit length of a vortex line, 
\begin{align}
  \bm{f}_{\rm  Mag}  &=  \rho \, (\delta  \bm{v})  \times  \bm{\kappa},
\end{align}
where $\rho$ is the superfluid density and $\bm{\kappa}$ is the vorticity vector, which is parallel to the vortex line (hence, parallel to the rotational axis of the NS) and has the absolute value $|\bm{\kappa}| \equiv \kappa$. 
\begin{figure}
\begin{center}
\centering
\includegraphics[width=0.8\textwidth]{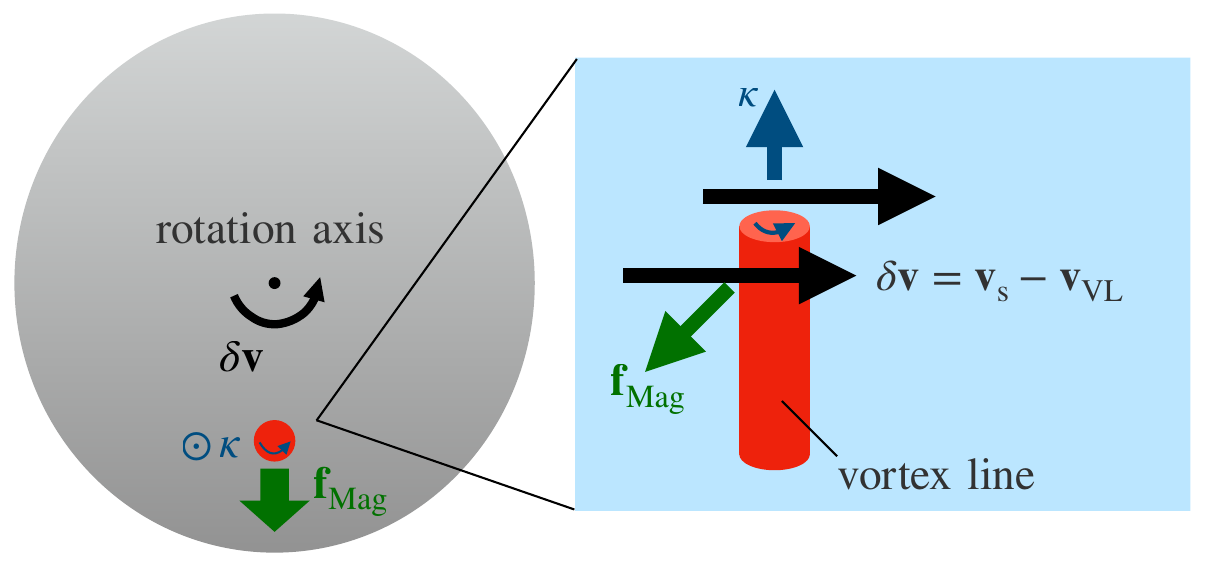}
\end{center}
\caption{The Magnus force acting on a vortex line. 
We define the direction of $\bm{\kappa}$ as the direction of the right-hand screw. 
  }
\label{fig:magnus}
\end{figure}
We see that the Magnus force always acts in the outward direction, as illustrated in Fig.~\ref{fig:magnus}, since the crust component rotates slower than the superfluid component due to the deceleration by the external torque. 

Vortex lines overcome the trapping through thermal fluctuations or quantum tunneling~\cite{1992PhyB..178....1B, 1993ApJ...403..285L}, and start to creep outward. If this creep rate is rapid enough, the superfluid rotation can smoothly follow the change in the crust rotation, and the system can reach a steady state. To see if this is the case for the NSs of our interest, let us briefly review the evaluation of the vortex creep rate $v_r$ following Ref.~\cite{1993ApJ...403..285L}. In this analysis, the pinning force is modeled by a periodic potential with a period equal to the span of the nuclei lattice $(\sim R_{\rm  WS})$ and a height equal to the pinning energy $(\sim  |E_{\rm pin}|)$. The Magnus force is considered as a bias that tilts the periodic potential. Let us introduce the transition rate for a vortex line to move from a local minimum into the next local minimum as ${\cal  R}_{\rm  VC}$. The zero-point frequency in the vicinity of local minima of the potential $\omega_0$ controls ${\cal  R}_{\rm  VC}$ and is obtained through the quantization of the vortex system~\cite{1992PhyB..178....1B, 1993ApJ...403..285L}. If the vortex tension is negligible compared to the pinning force, we obtain\footnote{In Ref.~\cite{1993ApJ...403..285L}, the case where the vortex tension dominates the pinning force is also studied, and the conclusion of the steady state turns out to remain unchanged.}
\begin{align}
    \omega_0 \simeq \frac{\pi \kappa \Lambda}{4 R_{\rm  WS}^2}\simeq 1.2\times10^{20}\,{\rm s}^{-1}
    \left( \frac{R_{\rm  WS}}{50~{\rm fm}} \right)^{-2}
    \left(\frac{\Lambda}{2}\right),
    \label{eq:omega}
\end{align}
where $\Lambda$ characterizes the vortex tension, $T_v = \rho \kappa^2\Lambda/(4\pi)$, and is evaluated as $2\lesssim \Lambda \lesssim 10$ in the inner crust~\cite{1991ApJ...373..592L}. The transition rate is then given by~\cite{1992PhyB..178....1B, 1993ApJ...403..285L}
\begin{align}
    {\cal  R}_{\rm  VC}  \simeq \frac{\omega_0}{2\pi}\, e^{-\frac{|E_{\rm  pin}|}{k_{\rm B} T_{\rm 
 eff}}},
 \label{eq: CreepRate}
\end{align}
where $T_{\rm  eff}$ is defined by 
\begin{align}
    k_{\rm B} T_{\rm  eff}  
    \equiv 
    \frac{\hbar  \omega_0}{2}\coth\left( 
    \frac{T_{\rm  q}}{T}
    \right)
\sim \begin{cases} k_{\rm B} T &(T  \gg  T_{\rm 
 q})\\
    \frac{\hbar \omega_0}{2}&(T  \ll  T_{\rm  q})
    \end{cases},
\end{align}
with  $T$ being the temperature of the inner crust and 
\begin{align}
  T_{\rm  q}  \equiv  \frac{\hbar \omega_0}{2k_{\rm B}}  \simeq  3.8  \times  10^8~\mathrm{K}  ~\left( 
 \frac{\omega_0}{10^{20}~\mathrm{s}^{-1}}\right).
\end{align}
As can be seen from these expressions, the unpinning occurs predominantly through thermal activation (quantum tunneling) for $T\gg T_{\rm  q}$ ($T\ll T_{\rm  q}$). In particular, for NSs as old as those considered in the following analysis, their internal temperature is $\lesssim 10^8$ K. Therefore, the vortex creep motion in these NSs is triggered by quantum tunneling. By using the transition rate in Eq.~\eqref{eq: CreepRate}, we evaluate the creep rate as $v_r\simeq {\cal  R}_{\rm  VC}  \cdot  R_{\rm WS}$. 

With the vortex creep rate obtained above, we can estimate the typical distance at which a vortex line travels through the creep motion during the lifetime of the NS: 
\begin{align}
    v_r\times t_{\rm age}
    &\simeq R_{\rm  WS}\cdot \frac{\omega_0}{2\pi}\cdot e^{-\frac{2|E_{\rm pin}|}{\hbar  \omega_0}}  \times  t_{\rm age}
    \nonumber
    \\
    &\simeq 160~{\rm km}\times \left(\frac{\omega_0}{10^{20}~\rm s^{-1}}\right)\left(\frac{R_{\rm  WS}}{50~\rm fm}\right)\left(\frac{t_{\rm age}}{10^5~\rm yr}\right),
    \label{eq:creep_distance}
\end{align}
where we set $|E_{\rm pin}|   =  1$~MeV as a representative value. We find that this distance is considerably larger than the crust thickness ($\simeq 1$~km), implying that the vortex creep motion would reach a steady state in old NSs. 

Once the system enters the steady phase, the crust and superfluid components decelerate at the same rate, i.e., 
\begin{align}
    \frac{\partial \Omega_{\rm  s} (t,r)}{\partial t}=\dot{\Omega}_{\rm  c} \equiv \dot{\Omega}_\infty~.
    \label{eq:steady_Omegadot}
\end{align}
Note that $\Omega_{\rm  c}$ and $|\dot{\Omega}_\infty|$ are identified as the current observed values of the angular velocity and deceleration rate of NSs, respectively. As a consequence, the relative angular velocity between the crust and superfluid components becomes independent of time, and this is found to be fairly close to the critical angular velocity determined by the condition $f_{\rm  pin}  =  f_{\rm  Mag}$~\cite{1993ApJ...403..285L}: 
\begin{align}
  \delta  \Omega_\infty  \simeq  \delta  \Omega_{\rm  cr}~,
\end{align}
with 
\begin{align}
  \delta  \Omega_{\rm  cr}  &\equiv  
  | \delta  \bm{\Omega} |_{f_{\rm  pin}  =  f_{\rm  Mag}}
  =  \frac{f_{\rm  pin}}{\rho  \kappa  r }.
  \label{eq:delta_Omega_cr}
\end{align}

\subsection{Prediction of surface temperature}
\label{sec:surface_temperature}

As vortices move outward, the rotational energy of the superfluid is dissipated through their frictional interaction with the normal components of the inner crust, which heats the NS. This heating luminosity is computed as~\cite{1984ApJ...276..325A,1989ApJ...346..808S}
\begin{align}
    L_{\mathrm{H}} &= 
    N_{\rm  ext}  \Omega_{\rm  c}  (t)  
  -  \frac{d}{dt}  
  \left[
    \frac{1}{2}  I_{\rm  c}  \Omega_{\rm  c}^2  (t)
    +
    \frac{1}{2}  \int  d  I_{\rm  p}  \Omega_{\rm  s}^2  (t,  r)
  \right] \nonumber \\ 
  &=  \int  d  I_{\rm  p} \left[ \Omega_{\rm  c}  (t) -  \Omega_{\rm  s}  (t,  r)  \right] \frac{\partial \Omega_{\rm  s} (t,r)}{\partial t} \nonumber \\ 
  &= \int  d  I_{\rm  p} \, \delta \Omega \left|\frac{\partial \Omega_{\rm  s} (t,r)}{\partial t}\right| ~,
\end{align}
where we use Eqs.~\eqref{eq:EoM_crust}~and~\eqref{eq:nint}, and the inertial momenta $I_{\rm  p}$ is integrated over the region where the pinning process efficiently occurs. This expression is further simplified in the steady state with the condition~\eqref{eq:steady_Omegadot},
\begin{align}
  L_{\mathrm{H}} &= J    | \dot{\Omega}_\infty | ~,
  \label{eq:Edot_def}
\end{align}
where we define,
\begin{align}
  J  \equiv  \int  d  I_{\rm  p}  \delta  \Omega_\infty ~,
  \label{eq:J_def}
\end{align}
and $\delta  \Omega_\infty$ denotes the steady-state value of the relative angular velocity. As we see, the heating luminosity is proportional to the current deceleration rate of the pulsar. Note that the value of $J$ can be estimated from Eqs.~\eqref{eq:steady_Omegadot}, \eqref{eq:delta_Omega_cr}, and \eqref{eq:J_def} by specifying the value of $f_{\rm  pin}$ and the region of pinning. We will evaluate the proportional coefficient $J$ in Sec.~\ref{sec:jeval}. 

In particular, if this heating luminosity balances with the photon luminosity, which we expect to occur for old NSs as we discussed in Sec.~\ref{sec:thermal_evolution}, the surface temperature $T_{\rm  s}$ can be estimated using Eq.~\eqref{eq:L_g=L_H} as 
\begin{align}
  T_{\rm  s}^{\rm  eq}  
  &\equiv
  \left( \frac{J  | \dot{\Omega}_\infty |}{4  \pi  R_{\rm  NS}^2  \sigma_{\rm SB}} \right)^{\frac{1}{4}}
  \\
  &\simeq
  1.0  \times  10^5~\mathrm{K}
  \left( \frac{J}{10^{43}~\mathrm{erg~s}} \right)^\frac{1}{4}
  \left( \frac{|\dot{\Omega}_\infty|}{10^{-14}~\mathrm{s}^{-2}} \right)^\frac{1}{4}
  \left( \frac{R_{\rm  NS}}{11.43~\mathrm{km}} \right)^{-\frac{1}{2}}.
  \label{eq:T_s_VC}
\end{align}
In the steady vortex creep scenario, therefore, the surface temperature is predicted as a function of $J$, $|\dot{\Omega}_{\infty}|$, and $R_{\rm  NS}$, free from the uncertainty of the initial condition and the subsequent temperature evolution. We will examine this prediction against observation in Sec.~\ref{sec:observation}.

\section{Theoretical approaches for the vortex pinning}
\label{sec:prediction}

To compute the energy dissipation due to vortex creep, we evaluate the parameter $J$ in Eq.~\eqref{eq:J_def}. We first review the calculation of the pinning force $f_{\rm  pin}$ available in the literature in Sec.~\ref{sec:fpineval}; the values of $f_{\rm  pin}$ which our analysis is based on are summarized in Appendix~\ref{sec:fpinvalues}. Then, in Sec.~\ref{sec:jeval}, we estimate possible ranges of $J$ using the results in Sec.~\ref{sec:fpineval}, which will be compared with observation in the subsequent section.

\subsection{Evaluation of pinning force}
\label{sec:fpineval}

To evaluate the pinning force, we need to analyze a nucleon many-body system at high densities, which generically suffers from technical difficulties due to less-known properties of nuclear interactions. A traditional method of treating nuclear interactions is to model a form of the interaction and fit it to experimental data, such as nucleon-nucleon scattering ~\cite{Hamada1962Nucl.Phys.34_382,Machleidt1989AdvancesinNuclearPhysicsVolume19_189,Wiringa:1994wb}. For example, the Argonne interaction~\cite{Wiringa:1994wb}, which we consider in the following analysis, is a two-body nucleon potential fitted to nucleon scattering data and deuteron properties. This sort of bare interaction does not include in-medium effects. The many-body calculation based on bare interaction is necessary to obtain the properties of nucleon systems. At the same time, it is still challenging to perform it in general due to, e.g., the strong repulsive core. An alternative method is to use an effective interaction incorporating in-medium effect phenomenologically. Skyrme-type interactions~\cite{Skyrme:1956zz, Skyrme1958Nucl.Phys.9_615} are well-known examples, which consist of contact (zero-range) interactions with momentum-dependent coefficients. The parameters of the interactions are determined by fitting to the experimental data of binding energies and radii of several nuclei. There are many sets of fitting parameters used in the literature~\cite{CHABANAT1997710,RevModPhys.75.121,PhysRevC.85.035201}, such as SLy4~\cite{Chabanat:1997un} and SkM*~\cite{BARTEL198279}. There are also finite-range interactions, such as the Gogny interaction~\cite{PhysRevC.21.1568}.

\begin{figure}
  \begin{center}
  \centering
  \includegraphics[width=1\textwidth]{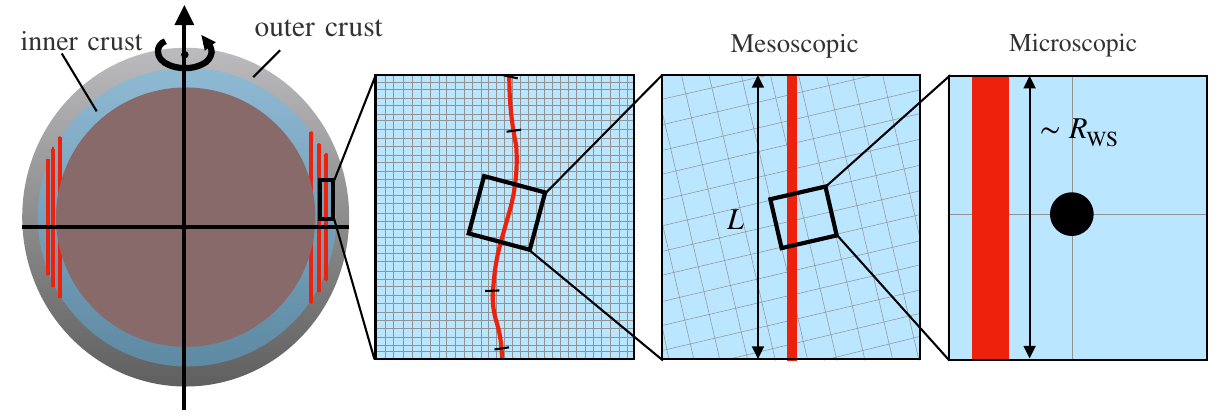}
  \end{center}
  \caption{The landscape of a vortex-line configuration at different length-scales. 
    }
  \label{fig:microscopic_vs_mesoscopic}
  \end{figure}

There are several approaches to analyzing the nuclear matter, and the following two are often used for the calculation of the pinning energy:
\begin{itemize}
  \item  \textbf{Quantum approach}
  
 A standard method to analyze a quantum multi-body system is to calculate the energy levels of a single particle in the mean field of a self-consistent potential by solving the corresponding Schr\"{o}dinger equation. A neutron pairing interaction is then considered to determine the pairing field as in the BCS theory. These solutions are obtained via an iterative process such that they satisfy the self-consistent conditions. This method is called the Hartree-Fock-Bogoliubov (HFB) method and adopted in Refs.~\cite{Avogadro:2006ed,Avogadro:2007kxn,Avogadro:2008uy,Klausner:2023ggr}.

  \item  \textbf{Semi-classical approach}

The quantum approach based on the HFB method often requires a high computational cost. To evade this, a semi-classical approach based on the Thomas-Fermi approximation is also frequently used, where nuclear matter is regarded as a many-body system of nucleons subject to the Pauli exclusion principle and moving independently from each other in a mean-field potential. The energy of the system for a given chemical potential is obtained by the variational principle.\footnote{Strictly speaking, we minimize the modified  Hamiltonian defined by $H^\prime = H - \mu N$, where $\mu$ and $N$ denote the chemical potential and the number of particles, respectively. We also note that for NSs, the temperature $T$ can be regarded as zero; thus, the free and internal energy are equivalent.} This approach is used in Refs~\cite{1988ApJ...328..680E,Pizzochero:1997iq,Donati:2003zz,Donati:2006mfa,Donati:2004gnw,Seveso:2016}. 
\end{itemize}
We can then estimate the pinning force from the pinning energy obtained above. There are two approaches for this calculation:  
\begin{itemize}
\item  \textbf{Microscopic calculation}

We may estimate the pinning force through Eq.~\eqref{eq:fpin_NP} for the nuclear pinning configuration. We refer to this estimation as the \textit{microscopic} approach since, as illustrated in the most right window in Fig.~\ref{fig:microscopic_vs_mesoscopic}, it focuses on the microscopic scale of $\mathcal{O}(R_{\rm  WS})$. This approach considers the interaction between a vortex and the single nucleus in the Wigner-Seitz cell, and thus the interaction of the vortex with other distant nuclei is neglected. We obtain the pinning force per unit length by just multiplying the pinning force per nucleus with the number of nuclei in the unit length along the vortex line ($\simeq 1/R_{\mathrm{WS}}$).

\item  \textbf{Mesoscopic calculation}

Vortex lines are much longer than the lattice spacing; therefore, each vortex line pins onto a large number of nuclei in reality. Such a vortex line does not align to the crystal axis over its total length in general. The \textit{mesoscopic} approach considers this realistic configuration by taking the average of the force exerted on a vortex over the possible directions of the vortex line with respect to the crystal lattice. This calculation focuses on the length-scale $L  \sim  (10^2 $--$ 10^3) \times R_{\rm  WS}$, for which the vortex line can be regarded as a straight line, as illustrated in the middle window in Fig.~\ref{fig:microscopic_vs_mesoscopic}---we call this scale the {mesoscopic} scale. The derived pinning force thus tends to be smaller than those obtained with the microscopic calculation.
\end{itemize}
\begin{table}
      \renewcommand{\arraystretch}{1.5}
      \centering
      \begin{tabular}{ c || c | c || c | c }
      \hline 
         &  \multicolumn{2}{c||}{\textbf{Semi-classical}} &   \multicolumn{2}{c}{\textbf{Quantum}}
        \\
        \hline 
        \hline
        \multirow{3}{*}{\textbf{Microscopic}\rule[-5mm]{0mm}{20mm}}
        & {\footnotesize Mean field:} & {\footnotesize Pairing:} & {\footnotesize Hartree-Fock:} & {\footnotesize Pairing:}
        \\
        \cline{2-5}
        & Woods-Saxon & Argonne
        \rule[-5mm]{0mm}{12mm}
        &  SLy4 & contact force  
        \\ 
        \cline{2-5}
        &
        \multicolumn{2}{c||}{$f_{\mathrm{pin}} = (1-7) \times 10^{-3}$~\cite{Donati:2004gnw}}
        &  
        \multicolumn{2}{c}{$f_{\mathrm{pin}} = (3-4) \times 10^{-4}$~\cite{Avogadro:2008uy}}
        \rule[-5mm]{0mm}{12mm}
        \\
        \hline
        \hline
          \multirow{3}{*}{\textbf{Mesoscopic}\rule[-5mm]{0mm}{20mm}}
        & {\footnotesize Mean field:} & {\footnotesize Pairing:} & {\footnotesize Hartree-Fock:} & {\footnotesize Pairing:}
        \\
        \cline{2-5}
        &  Woods-Saxon & Argonne
        &  SLy4 & contact force
        \rule[-5mm]{0mm}{12mm}
        \\
        \cline{2-5}
        &  \multicolumn{2}{c||}{$f_{\mathrm{pin}} = 8\times 10^{-7}-4 \times 10^{-4}$~\cite{Seveso:2016}}
        &
        \multicolumn{2}{c}{$f_{\mathrm{pin}} = 5 \times 10^{-7}-8 \times 10^{-5}$~\cite{Klausner:2023ggr}}
        \rule[-5mm]{0mm}{12mm}
        \\
        \hline
      \end{tabular}
      \caption{
        Calculations of the vortex pinning force considered in this work. The first row in each column lists the potentials for the mean-field and pairing interactions used in the evaluations; the second row shows the range of the pinning force in the inner crust in units of $\mathrm{MeV} \cdot \mathrm{fm}^{-2}$. 
      } 
      \label{tab:summary_main_refs}
      \end{table}
All in all, we have $2 \times 2 = 4$ combinations for the prescription of the pinning force calculation. In the following discussion, we consider a representative calculation with a specific choice of nuclear interactions for each combination, as summarized in Table~\ref{tab:summary_main_refs}. 

For the microscopic semi-classical approach, we consider the calculation in Ref.~\cite{Donati:2004gnw} with a Woods-Saxon potential for the mean-field potential and the Argonne interaction for the neutron-neutron pairing interaction. It is found that the nuclear pinning configuration (see Fig.~\ref{fig:configuration}) occurs only in high-density regions. The pinning force per nuclear pinning site is estimated by $E_{\mathrm{pin}}/R_{\mathrm{WS}}$ for this configuration in Ref.~\cite{Donati:2004gnw}. To convert this into the pinning force per unit length $f_{\mathrm{pin}}$, we multiply this by a factor of $1/(2 R_{\mathrm{WS}})$, as cubic cells with the side length $2 R_{\mathrm{WS}}$ are used in the calculation of Ref.~\cite{Donati:2004gnw}. As a result, we obtain the values of $f_{\mathrm{pin}} = \text{($1$--$7$)} \times 10^{-3}~\mathrm{MeV} \cdot \mathrm{fm}^{-2}$, depending on the position in the inner crust. We list quantities relevant to this calculation in Table~\ref{tab:vn-force_simulation_Micro-Argonne} in Appendix~\ref{sec:fpinvalues}.\footnote{To get some ideas about the dependence of this calculation on the choice of potentials, we note that Ref.~\cite{Donati:2004gnw} also considers the case where the Gogny interaction is used instead of the Argonne interaction. In this case, the nuclear pinning occurs in higher density regions, and the pinning forces tend to be larger by a factor of a few, compared with the calculation with the Argonne interaction; see Table~\ref{tab:vn-force_simulation_Micro-Gogny} in Appendix~\ref{sec:fpinvalues} for this calculation.  } Notice that the values of the pinning force shown here should be regarded as a ballpark estimate. The lower limit ($f_{\mathrm{pin}} = 1 \times 10^{-3}~\mathrm{MeV} \cdot \mathrm{fm}^{-2}$) could have been overestimated because of the discretization of the positions at which the pinning energy is estimated; as seen in Table~\ref{tab:vn-force_simulation_Micro-Argonne}, there is a density region of $\rho \sim 10^{13}~\mathrm{g} \cdot \mathrm{cm}^{-3}$ around which $E_{\mathrm{pin}} \simeq 0$, leading to a very small pinning force if we use Eq.~\eqref{eq:fpin_NP}. The upper limit could also be underestimated since the pinning force obtained by this equation corresponds to the average taken over the distance between the nuclear center and the interstitial position.

For the microscopic quantum approach, we consider the results given in Ref.~\cite{Avogadro:2008uy}, where the SLy4 Skyrme interaction is used for the Hartree-Fock calculation and a density-dependent contact interaction is used for the neutron-neutron pairing interaction. The parameters of the contact interaction are discussed in Ref.~\cite{Garrido:1999at}. Table~\ref{tab:vn-force_simulation_Micro-SLy4} summarises the relevant quantities for this calculation. In contrast to the semi-classical analysis~\cite{Donati:2004gnw}, nuclear pinning occurs in lower-density regions in this case. However, it also occurs in the highest-density regions (see Table~\ref{tab:vn-force_simulation_Micro-SLy4} in Appendix~\ref{sec:fpinvalues}).\footnote{The qualitative feature described here does depend on the choice of the nuclear interactions. As shown in Ref.~\cite{Avogadro:2008uy}, if we use the SkM* interaction instead of SLy4, the nuclear pinning never occurs at high densities. } The values of the pinning force estimated from the pinning energy and the Wigner-Seitz radius as in the semi-classical calculation are $f_{\mathrm{pin}} = \text{($3$--$4$)} \times 10^{-4}~\mathrm{MeV} \cdot \mathrm{fm}^{-2}$. See Ref.~\cite{Avogadro:2008uy} for detailed discussions regarding the difference between the semi-classical and quantum results. 

The semi-classical mesoscopic calculation is given in Ref.~\cite{Seveso:2016}, where the nuclear potentials are taken to be the same as in the semi-classical microscopic calculation in Table~\ref{tab:summary_main_refs}. The resultant values of the pinning force are found to be $f_{\mathrm{pin}} = 8\times 10^{-7}$ \text{--} $4 \times 10^{-4}~\mathrm{MeV} \cdot \mathrm{fm}^{-2}$, which are summarized in Table~\ref{tab:vn-force_SemiClassical-mesoscopic}. We see that these values are considerably smaller than those for the semi-classical microscopic calculation in Ref.~\cite{Donati:2004gnw} due to the averaging over the vortex-line directions. 

For the quantum mesoscopic calculation, we consider the result given in Ref.~\cite{Klausner:2023ggr}, where the SLy4 interaction and a contact interaction are used for the Hartree-Fock calculation and pairing interactions, respectively, as in the quantum microscopic calculation in Table~\ref{tab:summary_main_refs}. As summarized in Table~\ref{tab:vn-force_Quantum-mesoscopic}, the pinning force is found to be in the range $f_{\mathrm{pin}} = 5 \times 10^{-7}-8 \times 10^{-5}~\mathrm{MeV} \cdot \mathrm{fm}^{-2}$.\footnote{If we use the SkM* potential instead of SLy4, we obtain slightly smaller values of $f_{\mathrm{pin}}$, as also shown in Table~\ref{tab:vn-force_Quantum-mesoscopic}. } We again find that these values are much smaller than those obtained with the quantum microscopic approach. 

Before concluding this section, we note that we can also calculate the pinning force with a three-dimensional dynamical simulation of a vortex~\cite{Link:2008aq, Bulgac:2013nmn, Wlazlowski:2016yoe, Link:2022vum}.\footnote{See also Ref.~\cite{Antonelli:2020umo} for a kinetic approach.} Such a calculation tends to be costly; thus, a certain degree of simplification is usually required for the moment. 
Besides, the current evaluation is limited to a few benchmark values of density. 
The estimated values of the pinning force are consistent with the above estimates.

\subsection{Theoretical evaluation of $J$}
\label{sec:jeval}

For the evaluation of $J$ in Eq.~\eqref{eq:J_def}, it is convenient to change the coordinate from cylindrical coordinates $( r, \varphi, z )$ to spherical coordinates $( R, \theta,  \phi )$:
\begin{align}
  J
  &=
  \int_{\rm  pin}  d  r  \,  d  z  \,  d  \varphi  \,  \rho r^3  \cdot \frac{f_{\mathrm{pin}}}{\rho \kappa r}
  \simeq
  \int_{R_{\rm  in}}^{R_{\rm  out}}  d  R  \,  d  \theta  \,  d  \phi  \,  R^3  \sin^2  \theta
  \cdot  \frac{f_{\rm  pin}}{\kappa}\,,
  \label{eq:J_approx}
\end{align}
where we approximate $\delta \Omega_{\infty}$ in Eq.~\eqref{eq:J_def} by $\delta \Omega_{\mathrm{cr}}$ in Eq.~\eqref{eq:delta_Omega_cr}, which holds with good accuracy in the situation of our interest~\cite{1993ApJ...403..285L} as we mentioned in Sec.~\ref{sec:dynamics}. We perform the integral over the range  $[ R_{\rm  in}, R_{\rm  out} ]$  where the pinning force is evaluated. We use the Akmal-Pandharipande-Ravenhall (APR)~\cite{Akmal:1998cf} equation of state to determine the NS core size and the equation of state tabulated in \texttt{Crust\_EOS\_Cat\_HZD-NV.dat} in  \texttt{NSCool}~\cite{NSCool} based on Refs.~\cite{Negele:1971vb, 1989A&A...222..353H} to determine the density distribution in the crust. 

For the evaluation method of pinning force, we focus on the mesoscopic approach shown in Table~\ref{tab:summary_main_refs}, since for the microscopic calculation, the pinning force is obtained only in a limited region in the crust, as can be seen in Table~\ref{tab:vn-force_simulation_Micro-Argonne}--\ref{tab:vn-force_simulation_Micro-SLy4}. Considering that the evaluation of the pinning force suffers from large uncertainty depending on calculation methods, we make the following crude approximation in the calculation of the above integral---we neglect the density dependence of $f_{\mathrm{pin}}$ and fix it to a value in the range shown in Table~\ref{tab:summary_main_refs}. We thus obtain a range of $J$ accordingly, which we regard as the uncertainty of this pinning force estimation. As a result, we obtain $J = 3.9  \times  10^{40} - 1.9  \times  10^{43}~\mathrm{erg} \cdot \mathrm{s}$ for the semi-classic mesoscopic calculation and $J = 1.7  \times  10^{40} - 2.7  \times  10^{42} ~\mathrm{erg} \cdot \mathrm{s}$ for the quantum mesoscopic calculation.

\section{Vortex creep heating vs. observation}
\label{sec:observation}

{\footnotesize
    \begin{table}
        \renewcommand{\arraystretch}{1.5}
        \centering
        \scalebox{0.8}{
        \begin{tabular}{lll  ||  ll  |  ll  |  l}
          \hline
          No.
          &  Type
          &  Name
          &  $\log_{10}  t_{\rm  sd}$  
          &  $\log_{10}  t_{\rm  kin}$    
          &  $\log_{10}  | \dot{\Omega} |$  
          &  $\log_{10}  T_{\rm  s}$  
          &  $\log_{10}  J_{\rm  obs}$
          \\
          &
          &
          &  [yr]  
          &  [yr]  
          &  [s$^{-2}$]
          &  [K]  
          &  [erg~s]
          \\
          \hline
          \hline
          1.
          &  \textcolor{gray}{[Y]}
          &  PSR B1706-44  
          &  $4.2$  &  ---    
          &  $-10.3$
          &  $5.68$--$6.34$~\cite{McGowan:2003sy}  
          &  $41.9$--$44.6$
          \\
          2.
          &  \textcolor{blue}{[O]}
          &  PSR J1740+1000
          &  $5.1$  
          &  ---
          &  $-11.2$
          &  $5.89$~\cite{Kargaltsev:2012yi}
          &  $43.8$
          \\         
          3.
          &  \textcolor{gray}{[Y]}
          &  PSR B2334+61
          &  $4.6$  &  ---
          &  $-11.3$
          &  $5.76$~\cite{McGowan:2005kt}
          &  $43.3$
          \\
          4.
          &  \textcolor{blue}{[O]}
          &  PSR B0656+14
          &  $5.0$  
          &  ---
          &  $-11.6$
          &  $5.87$~\cite{Arumugasamy:2018pli}
          &  $44.1$
          \\
          5.
          &  \textcolor{blue}{[O]}
          &  PSR J0633+1746
          &  $5.5$  
          &  ---
          &  $-11.9$
          &  $5.71$~\cite{Mori:2014gaa}
          &  $43.7$
          \\   
          6.
          &  \textcolor{gray}{[Y]}
          &  PSR J0538+2817
          &  $5.8$  
          &  $4.60^{+0.18}_{-0.30}$\cite{Ng:2006vh}  
          &  $-11.9$
          &  $6.02$~\cite{Ng:2006vh}
          &  $45.0$
          \\
          7.
          &  \textcolor{blue}{[O]}
          &PSR B1055-52
          &  $5.7$  
          &  ---
          &  $-12.0$
          &  $5.81$~\cite{DeLuca:2004ck}
          &  $45.1$
          \\
          8.
          &  \textcolor{OliveGreen}{[X]}
          &  RX J1605.3+3249
          &  $4.5$  
          &  $5.66^{+0.04}_{-0.07}$~\cite{Tetzlaff:2012rz}
          &  $-12.1$
          &  $5.86$~\cite{Malacaria:2019zqr}
          &  $44.5$
          \\
          9.
          &  \textcolor{blue}{[O]}
          &  PSR J2043+2740
          &  $6.1$  
          &  ---
          &  $-12.1$
          &  $<  5.95$~\cite{Beloin:2016zop}
          &  $<  44.8$
          \\
          10.
          &  \textcolor{blue}{[O]}
          &  PSR J1741-2054
          &  $5.6$  
          &  ---
          &  $-12.2$
          &  $5.85$~\cite{Auchettl:2015wca}
          &  $44.6$
          \\
          11.
          &  \textcolor{blue}{[O]}
          &  PSR J0357+3205
          &  $5.7$  
          &  ---
          &  $-12.4$
          &  $5.62$~\cite{Kirichenko:2014ona}
          &  $43.8$
          \\
          12.
          &  \textcolor{blue}{[O]}
          &  PSR B0950+08
          &  $7.2$  
          &  ---
          &  $-13.6$
          &  $4.78$--$5.08$~\cite{Abramkin:2021fzy}
          &  $41.7$--$42.9$
          \\
          13.
          &  \textcolor{OliveGreen}{[X]}
          &  RX J0420.0-5022
          &  $6.3$  
          &  ---
          &  $-13.8$
          &  $5.74$~\cite{Kaplan:2011xd}
          &  $45.8$
          \\
          14.
          &  \textcolor{Orange}{[M]}
          &  PSR J0437-4715
          & $9.20$
          &  ---
          &  $-14.0$
          &  $5.54$~\cite{Durant:2011je}
          &  $45.1$
          \\
          15.
          &  \textcolor{OliveGreen}{[X]}
          &  RX J1308.6+2127
          &  $6.2$  
          &  $5.74^{+0.16}_{-0.26}$~\cite{Motch:2009nq}
          &  $-14.2$
          &  $6.08$~\cite{Schwope:2006ra}
          &  $47.5$
          \\        
          16.
          &  \textcolor{OliveGreen}{[X]}
          &  RX J0720.4-3125
          &  $6.3$  
          &  $5.93^{+0.07}_{-0.26}$~\cite{Tetzlaff:2011kh} 
          &  $-14.2$
          &  $6.02$~\cite{Hambaryan:2017wvm}
          &  $47.3$
          \\
          17.
          &  \textcolor{Orange}{[M]}
          &  PSR J2124-3358
          &  $9.58$
          &  ---
          &  $-14.3$
          &  $4.70$--$5.32$~\cite{Rangelov:2016syg}
          &  $42.0$--$44.5$
          \\
          18.
          &  \textcolor{OliveGreen}{[X]}
          &  RX J1856.5-3754
          &  $6.6$  
          &  $5.66^{+0.04}_{-0.05}$\cite{Tetzlaff:2011kh}
          &  $-14.4$
          &  $5.65$~\cite{Sartore:2012fk}
          &  $46.0$
          \\
          19.
          &  \textcolor{OliveGreen}{[X]}
          &  RX J2143.0+0654
          &  $6.6$  
          &  ---
          &  $-14.5$
          &  $5.67$--$6.06$~\cite{Kaplan:2009au,Schwope:2009tv}
          &  $46.1$--$47.8$
          \\
          20.
          &  \textcolor{OliveGreen}{[X]}
          &  RX J0806.4-4123
          &  $6.5$  
          &  ---
          &  $-14.6$
          &  $6.01$~\cite{Kaplan:2009ce}
          &  $47.6$
          \\
          21.
          &  \textcolor{blue}{[O]}
          &  PSR J0108-1431
          &  $8.3$  
          &  ---
          &  $-15.1$
          &  $4.43$--$4.74$~\cite{Abramkin:2021tha}
          &  $41.8$--$43.1$
          \\
          22.
          &  \textcolor{blue}{[O]}
          &  PSR J2144-3933
          &  $8.4$  
          &  ---
          &  $-16.4$
          &  $< 4.62$~\cite{Guillot:2019ugf}
          &  $< 43.8$
          \\
          \hline
        \end{tabular}
        }
        \caption{
          The data of the NSs considered in this paper. We classify them into four types---ordinary pulsars younger than $10^5~\mathrm{yrs}$ \textcolor{gray}{[Y]}, ordinary pulsars older than $10^5~\mathrm{yrs}$ \textcolor{blue}{[O]}, 
          XDINSs \textcolor{OliveGreen}{[X]}, 
          and millisecond pulsars \textcolor{Orange}{[M]}. The values of $t_{\rm  sd}$ and $\dot{\Omega}$ without references are computed from the data given in the ATNF pulsar catalogue~\cite{Manchester:2004bp, ATNF_catalogue}. The value of $J_{\rm  obs}$ is evaluated as in Eq.~\eqref{eq:J_obs} with $R_{\rm  NS} = 11.43~\mathrm{km}$. 
          } 
        \label{tab:plusar_catalogue}
    \end{table}
}

We now compare the prediction of the vortex-creep heating mechanism with observation. For this purpose, it is useful to calculate the following quantity for each NS:\footnote{We neglect the gravitational redshift factor since its effect is within the $\mathcal{O}(1)$ uncertainty of $J$ discussed below.} 
\begin{align}
  J_{\rm  obs}  &\equiv  \frac{4\pi R_{\rm NS}^2\sigma_{\rm  SB}T_{\rm  s}^4}{|\dot{\Omega}|} ~. 
  \label{eq:J_obs}
\end{align}
As evident from Eqs.\eqref{eq:Lgamma_eq_LH}, \eqref{eq:L_g=L_H}, and \eqref{eq:Edot_def}, this corresponds to the $J$ parameter inferred from the observation of each NS. This inference assumes the steady creeping of vortices (discussed in Sec.\ref{sec:dynamics}) and the balance between vortex-creep heating luminosity and photon cooling luminosity, which we expect to hold if the NS is older than $\sim 10^5$~years. Since NSs are comparable in size and mass, we expect that $J_{\rm  obs}$ is also roughly equal (up to a factor of $\mathcal{O}(1)$) for every NS. We test this expectation by using the data of $J_{\rm  obs}$ for old NSs. We also compare the values of $J_{\mathrm{obs}}$ with the theoretical computations given in Sec.~\ref{sec:jeval}. 

In Table~\ref{tab:plusar_catalogue}, we list the values of $J_{\mathrm{obs}}$ for the NSs we consider in this paper. We select isolated NSs older than $10^4$~yrs. In evaluating $J_{\mathrm{obs}}$, we have just assumed $R_{\rm  NS} = 11.43~\mathrm{km}$ for all NSs, as the radius is poorly known for most NSs; we keep in mind that this may introduce an $\mathcal{O}(1)$ error in the determination of $J_{\mathrm{obs}}$. We also show the age, surface temperature, and $\dot{\Omega} = 2  \pi  \dot{P}/P^2$ of the NSs, ($P$ and $\dot{P}$ are the period and its time derivative, respectively). Regarding the NS age, we use the kinetic age $t_{\rm  kin}$ if available. Otherwise, we use the spin-down age $t_{\rm  sd}  =  P/(2\dot{P})$. We calculate $t_{\rm  sd}$ and $\dot{\Omega}$ from $P$ and $\dot{P}$ given in the Australia Telescope National Facility (ATNF) pulsar catalogue~\cite{Manchester:2004bp,ATNF_catalogue}. Notice that the surface temperatures of some of the old NSs in this table are much higher than the predicted temperature in the standard NS cooling scenario~\cite{Yakovlev:1999sk,Yakovlev:2000jp,Yakovlev:2004iq,Page:2004fy,Gusakov:2004se,Page:2009fu}. 

It is important to note that not all NSs listed in Table~\ref{tab:plusar_catalogue} are useful for testing the vortex-creep heating. As we have discussed in Sec.~\ref{sec:thermal_evolution}, photon emission becomes the dominant cooling source for NSs older than $\sim 10^5$ years. For younger NSs, this may not be the case, so $L^\infty_\gamma \lesssim L^\infty_{\rm  H}$ instead of \eqref{eq:L_g=L_H}, for which the values of $J_{\mathrm{obs}}$ in Table~\ref{tab:plusar_catalogue} may be underestimated. To distinguish such young NSs from others, we indicate them by the type \textcolor{gray}{[Y]} in the table. Another class of NSs that are inappropriate for our test is the X-ray Dim Isolated NSs (XDINSs). These NSs are considered to be descendants of magnetars~\cite{Kaplan:2009ce, Kaplan:2011xd} that experienced the decay of magnetic fields before. This may make these NSs hotter than ordinary NSs of the same age~\cite{Pons:2008fd, Kaplan:2011xd, 2012MNRAS.422.2878D, Vigano:2013lea}, resulting in an overestimate of $J_{\mathrm{obs}}$.  We denote these NSs by the type \textcolor{OliveGreen}{[X]}. The rest of the NSs, which we use for the test of the vortex-creep heating, are classified into old ordinary pulsars \textcolor{blue}{[O]} and millisecond pulsars \textcolor{Orange}{[M]}.

The uncertainty in the determination of $J_{\mathrm{obs}}$ stems mainly from that in the surface temperature, which is significant due to its quartic dependence on $T_{\rm  s}$. Generically speaking, it is very difficult to identify all of the sources of uncertainties in the measurement of the NS surface temperature, and it is often the case that the error shown in the literature is only a part of them, such as those from the spectrum fitting, the determination of the distance and/or radius of the NS, and so on. At present, it is fair to say that the NS temperature measurement typically suffers from $\mathcal{O} (1)$ uncertainty, as can be seen in, \textit{e.g.}, Ref.~\cite{Potekhin:2020ttj}. Motivated by this, we include a factor of two uncertainty in $T_{\rm  s}$ for the stars in Table~\ref{tab:plusar_catalogue} for which only the central value is presented. For the other stars, we describe our prescription for the error estimation in Appendix~\ref{sec:catalog_selection}. We have checked that the errors thus obtained are similar to or more conservative than those adopted in Ref.~\cite{Potekhin:2020ttj}. 

\begin{figure}
  \begin{center}
\includegraphics[width=1\textwidth]{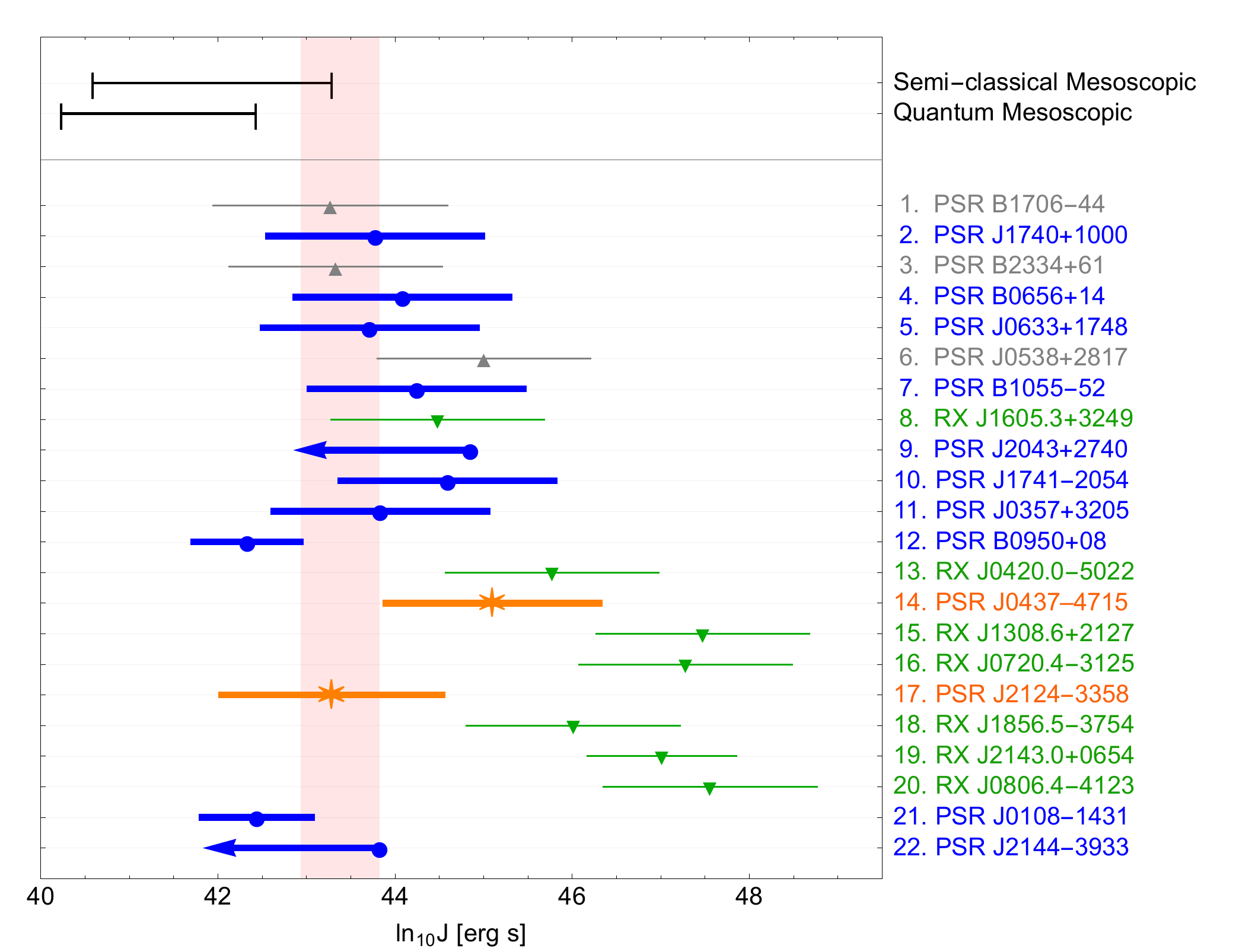}
  \end{center}
\caption{
The values of the $J$ parameter obtained from the observation. 
The grey triangles, 
blue circles, 
green inverse triangles, 
and orange stars
correspond to the young ordinary pulsars (\textcolor{gray}{[Y]}), 
the old ordinary pulsars (\textcolor{blue}{[O]}), 
the XDINSs (\textcolor{OliveGreen}{[X]}), and 
the millisecond pulsars (\textcolor{Orange}{[M]}), respectively. 
The points with an arrow indicate upper limits. The red shaded region shows observationally favored range, $J \simeq 10^{42.9\text{--}43.8} ~\mathrm{erg} \cdot \mathrm{s}$. For comparison, we also show the values of $J$ estimated with the mesoscopic calculations by black bars.
}
\label{fig:Jpin_vs_Jobs}
\end{figure}

In Fig.~\ref{fig:Jpin_vs_Jobs}, we show the range of $J_{\mathrm{obs}}$ estimated as described above for each NS listed in Table~\ref{tab:plusar_catalogue}. 
The grey triangles, green inverse triangles, blue circles, 
and orange stars represent the young ordinary pulsars (\textcolor{gray}{[Y]}), the XDINSs (\textcolor{OliveGreen}{[X]}), the old ordinary pulsars (\textcolor{blue}{[O]}), 
and the millisecond pulsars (\textcolor{Orange}{[M]}), respectively. 
The points with an arrow indicate that we only have an upper limit on $J_{\rm  obs}$ for those NSs. Recall that we are concerned only with the NSs represented by the blue \textcolor{blue}{[O]} and orange \textcolor{Orange}{[M]} points. It is found that the estimated values of $J_{\mathrm{obs}}$ for these NSs are in the same ballpark, $J \sim 10^{43}~\mathrm{erg} \cdot \mathrm{s}$, even though their $|\dot{\Omega}|$'s distribute over orders of magnitude. This is in good agreement with the prediction of the vortex-creep heating mechanism. 
On the other hand, $J_{\mathrm{obs}}$ for the green points (XDINSs \textcolor{OliveGreen}{[X]}) tend to be larger than this, as expected.

We also show the theoretical estimations given in Sec.~\ref{sec:jeval} in the upper panel of Fig.~\ref{fig:Jpin_vs_Jobs}. We see that the semi-classical mesoscopic calculation is consistent with the observation, given that this theoretical estimation suffers from a NS-dependent uncertainty of $\mathcal{O}(1)$ coming from the integration in Eq.~\eqref{eq:J_approx}, in addition to that from the estimation of $f_{\mathrm{pin}}$. The quantum mesoscopic calculation can explain some of the points with a small $J_{\mathrm{obs}}$, but they are not large enough to explain, \textit{e.g.}, that of J0437-4715. However, we note that this theoretical estimation is still allowed by the observations since it just results in a lower heating luminosity than the observed one. If this is the case, the vortex-creep heating may operate but there exists another heating mechanism that dominates the vortex-creep heating, such as the rotochemical heating~\cite{Reisenegger:1994be, 1992A&A...262..131H, 1993A&A...271..187G, Fernandez:2005cg, Villain:2005ns, Petrovich:2009yh, Pi:2009eq, Gonzalez-Jimenez:2014iia, Yanagi:2019vrr}. 

It is premature to establish the existence of the vortex-creep heating, as well as to conclude if an extra heating mechanism is required to be present. To that end, we need to accumulate more data on the surface temperature of old NSs with high accuracy, which we anticipate to be provided by future optical, UV, and X-ray observations.\footnote{See, for instance, Ref.~\cite{Toyouchi:2021ajr}.} Nevertheless, obtaining a current compilation of the value of $J$ suggested by the observation is intriguing. Considering intrinsic $\mathcal{O} (1)$ uncertainty in $J$, we determine its rough range by requiring that it covers the range suggested by B0950+08, which favors the smallest value, and satisfies the upper limit set by J2144-3944 based on non-observation of thermal flux. This yields 
\begin{align}
  J \simeq 10^{42.9 - 43.8} ~\mathrm{erg} \cdot \mathrm{s} ~,
  \label{eq:jrange}
\end{align}
which we show as the red band in Fig.~\ref{fig:Jpin_vs_Jobs}.

\begin{figure}
  \centering
  \subcaptionbox{\label{fig:teff_o}
  Ordinary pulsars  
  }
  {\includegraphics[width=0.47\textwidth]{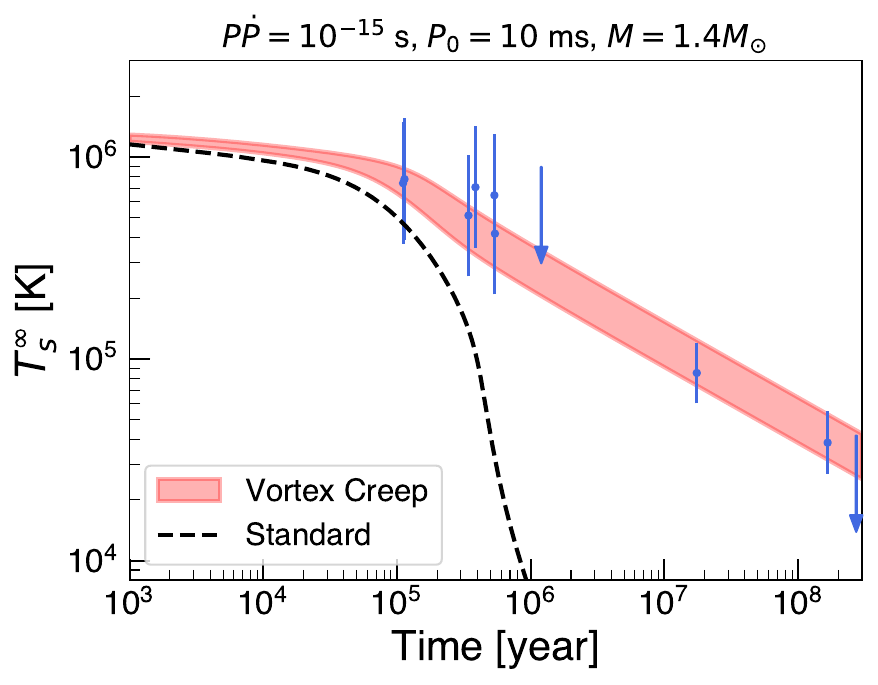}}
  \subcaptionbox{\label{fig:teff_m}
  Millisecond pulsars 
  }
  { 
  \includegraphics[width=0.48\textwidth]{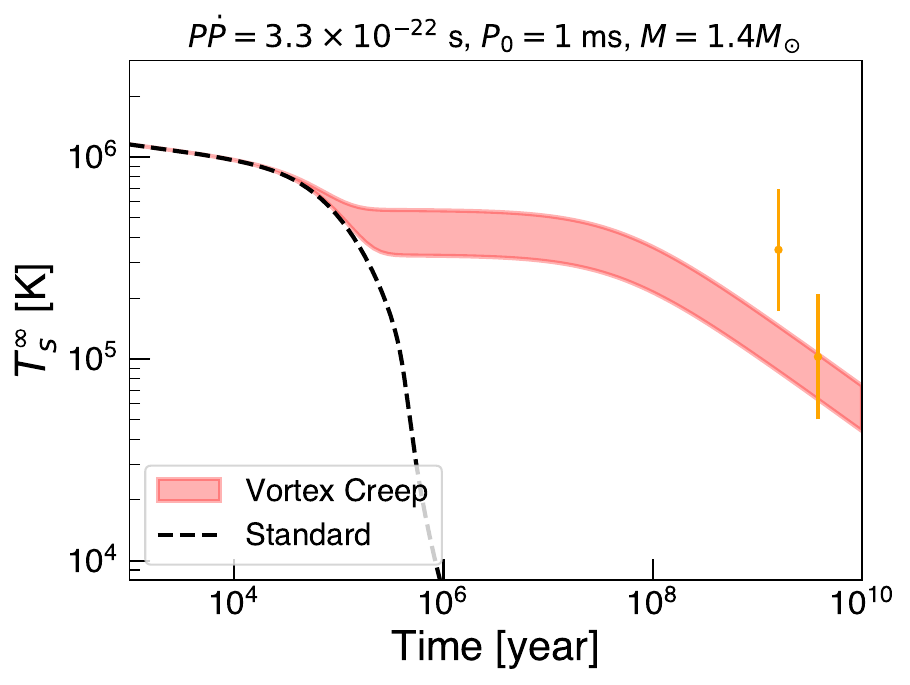}}
  \caption{The evolution of NS surface temperature with (without) the vortex creep heating effect in the red band (black dashed line). The band corresponds to the range of $J$ in Eq.~\eqref{eq:jrange}. The dots with error bars show the observed temperatures, presented in Table~\ref{tab:plusar_catalogue}, with the same colors as in Fig.~\ref{fig:Jpin_vs_Jobs}. 
  }
  \label{fig:teff}
  \end{figure}

Finally, in Fig.~\ref{fig:teff}, we show the evolution of NS surface temperature with (without) the vortex creep heating effect in the red band (black dashed line).\footnote{In these plots, we use the APR equation of state~\cite{Akmal:1998cf} for a NS mass of $1.4M_\odot$ to calculate the NS structure. For Cooper pairing gap models, we use the SFB model~\cite{Schwenk:2002fq} for the neutron singlet pairing, the model ``b'' in Ref.~\cite{Page:2004fy} for the neutron triplet pairing, and the CCDK model~\cite{Chen:1993bam} for the proton singlet pairing. The temperature evolution at late times scarcely depends on the choice of these models.  } The band corresponds to the range of $J$ in Eq.~\eqref{eq:jrange}. The dots with error bars show the observed temperatures in Table~\ref{tab:plusar_catalogue}, with the same colors as in Fig.~\ref{fig:Jpin_vs_Jobs}. In Fig.~\ref{fig:teff_o}, we take $P\dot{P} = 10^{-15}$~s and the initial period $P_0 = 10~\mathrm{ms}$ to calculate $|\dot{\Omega} (t)|$, where we assume that the external torque is dominated by magnetic dipole radiation.\footnote{In this case, we have $\dot{\Omega} \propto - \Omega^3$, i.e., $P\dot{P} = \text{constant}$, and by solving this we obtain 
\begin{align}
  |\dot{\Omega}|(t) =  \frac{\pi}{\sqrt{2P \dot{P}}} \left[t + t_{\mathrm{sd},0}\right]^{-3/2} ,
  \nn
\end{align}
with $t_{\mathrm{sd},0} \equiv P_0^2 /(2 P \dot{P})$. For the choice of parameters in Fig.~\ref{fig:teff_o} and Fig.~\ref{fig:teff_m}, $t_{\mathrm{sd},0} \simeq  2 \times 10^3$ and $5 \times 10^7$~years, respectively. The value of $P \dot{P}$ is related to the surface magnetic flux density $B_s$. In the ATNF pulsar catalogue~\cite{Manchester:2004bp, ATNF_catalogue}, $B_s = 3.2 \times 10^{19} (P \dot{P})^{1/2}$~G is used for this relation, with which we have $B_s \simeq 1.0\times 10^{12}$~G and $5.8 \times 10^8$~G in Fig.~\ref{fig:teff_o} and Fig.~\ref{fig:teff_m}, respectively.} These values are typical for ordinary pulsars. Nevertheless, we note that $|\dot{\Omega} (t)|$ obtained with these parameters do not exactly agree with the observed values of $|\dot{\Omega}|$ in Table~\ref{tab:plusar_catalogue}, so the data points shown in this figure should be regarded as just an eye guide. In Fig.~\ref{fig:teff_m}, we set $P\dot{P} = 3.3 \times 10^{-22}$~s, which is the observed value for PSR J2124-3358, and $P_0 = 1~\mathrm{ms}$. As we see in these plots, the predicted temperature with the vortex creep heating starts to deviate from that in the standard cooling scenario at $t \sim 10^5$~years and remains high enough at later times to be compatible with the observed data.

\section{Conclusion and discussion}
\label{sec:conclusion}

We have revisited the vortex-creep heating mechanism in light of recent observations of old warm NSs. As we have seen, this heating mechanism gives a characteristic prediction that the heating luminosity is proportional to $|\dot{\Omega}|$, with the proportional constant $J$ having an almost universal value over NSs since the NS structure and the vortex-nuclear interactions determine it. We have found that this prediction agrees with the observational data of old NSs, with the favored range of $J$ in the same ballpark as the theoretical calculations. 

Notice that the scenario where vortex creep heating dominates all NSs can readily be overturned if we discover a NS having $J$ much smaller than those presented in Fig.~\ref{fig:Jpin_vs_Jobs}. On the other hand, if we find a NS with a larger $J$, we can disfavor our scenario only after excluding the existence of other heating sources specific to this NS, such as accretion from its environment. 

It is possible that other heating mechanisms also work in old NSs. Indeed, we have already considered potential heating caused by the decay of magnetic fields in XDINSs (see Sec.~\ref{sec:observation}), and we have not used these NSs in our test of the vortex-creep heating mechanism for this reason. Another heating mechanism that may operate without relying on exotic phenomena is provided by the out-of-equilibrium beta processes, which is dubbed rotochemical heating~\cite{Reisenegger:1994be, 1992A&A...262..131H, 1993A&A...271..187G, Fernandez:2005cg, Villain:2005ns, Petrovich:2009yh, Pi:2009eq, Gonzalez-Jimenez:2014iia, Yanagi:2019vrr}. It is known that this rotochemical heating mechanism can increase the surface temperature of old NSs up to $\sim 10^6$~K. Thus, its heating luminosity can be comparable to or even dominate the vortex-creep heating. It would be worthwhile to study the vortex creep heating in the presence of rotochemical heating and compare its prediction with the temperature observations of old warm NSs.   The NS heating caused by the accretion of dark matter particles is also widely discussed in the literature~\cite{Kouvaris:2007ay, Bertone:2007ae, Kouvaris:2010vv, deLavallaz:2010wp, Bramante:2017xlb, Baryakhtar:2017dbj, Raj:2017wrv,  Chen:2018ohx, Bell:2018pkk, Garani:2018kkd, Camargo:2019wou, Bell:2019pyc, Hamaguchi:2019oev, Garani:2019fpa, Acevedo:2019agu, Joglekar:2019vzy, Keung:2020teb,  Yanagi:2020yvg, Joglekar:2020liw, Bell:2020jou, Bell:2020lmm, Anzuini:2021lnv, Zeng:2021moz, Bramante:2021dyx, Tinyakov:2021lnt,Maity:2021fxw, Fujiwara:2022uiq, Hamaguchi:2022wpz, Chatterjee:2022dhp, Coffey:2022eav, Acuna:2022ouv, Alvarez:2023fjj, Bramante:2023djs}. In this case, the surface temperature at late times is predicted to be $\text{a few} \times 10^3$~K, and thus this heating effect could be concealed by the vortex heating mechanism, making it improbable to observe the dark matter signature through the temperature observation of old NSs. A detailed study of this issue will be given in the forthcoming paper~\cite{FHNRpaper2}.

\section*{Acknowledgments}
\noindent
MF thanks Kazuyuki Sekizawa and Tomoya Naito for the fruitful discussion about the current situation and the possible future directions in the pinning force evaluation. 
This work is supported in part by the Collaborative Research Center SFB1258 and by the Deutsche Forschungsgemeinschaft (DFG, German Research Foundation) under Germany's Excellence Strategy - EXC-2094 - 390783311 [MF], JSPS Core-to-Core Program (No. JPJSCCA 20200002 [MF]), the Grant-in-Aid for Innovative Areas (No.19H05810 [KH and MRQ], No.19H05802 [KH], No.18H05542 [MF and NN]), Scientific Research B (No.20H01897 [KH and NN]), Young Scientists (No.21K13916 [NN]).

\appendix

\section{Pinning force}
\label{sec:fpinvalues}

In this appendix, we show the density dependence of the quantities relevant to the pinning force calculations discussed in Sec.~\ref{sec:fpineval}. Tables~\ref{tab:vn-force_simulation_Micro-Argonne} and \ref{tab:vn-force_simulation_Micro-Gogny} are for the microscopic semi-classical approach in Ref.~\cite{Donati:2004gnw}, where the Argonne and Gogny interactions are used for the nuclear pairing interaction, respectively. The element corresponding to the cell nuclear composition and Wigner-Seitz radius $R_{\rm  WS}$ are derived for each baryon density $\rho$ in Ref.~\cite{Negele:1971vb}. In Ref.~\cite{Donati:2004gnw}, the pinning force is evaluated only for the nuclear pinning configuration, and thus we show the values of $E_{\mathrm{pin}}$ and $f_{\mathrm{pin}} = |E_{\mathrm{pin}}|/(2 R_{\mathrm{WS}}^2)$ only for this case. The labels of NP and IP in the last column indicate the nuclear and interstitial pinnings, respectively. 

\begin{table}
  \renewcommand{\arraystretch}{1.5}
  \centering
  \begin{tabular}{ c || c | c | c  c  c | c | c }
    \hline
    zone  &  Element  &  $\rho$ &  $R_{\rm  WS}$  &  $\xi$  &  $E_{\rm  pin}$  &  $f_{\rm  pin}$  &  config
    \\
    &&  [g  cm$^{-3}$]  &  [fm]  &  [fm]  &  [MeV]  &  [MeV  fm$^{-2}$]  &
    \\
    \hline
    \hline
    1  &  $^{320}_{40}\mathrm{Zr}$  &  $1.5  \times  10^{12}$  &  $44.0$  &  $7.02$  &  ---  &  ---  &  IP
    \\
    2  &  $^{1100}_{50}\mathrm{Sn}$  &  $9.6  \times  10^{12}$  &  $35.5$  &  $4.34$  &  ---  &  ---  &  IP
    \\
    3  &  $^{1800}_{50}\mathrm{Sn}$  &  $3.4  \times  10^{13}$  &  $27.0$  &  $8.54$  &  $-5.2$  &  $0.0036$  &  NP
    \\
    4  &  $^{1500}_{40}\mathrm{Zr}$  &  $7.8  \times  10^{13}$  &  $19.4$  &  $11.71$  &  $-5.1$  &  $0.0068$  &  NP
    \\  
    5  &  $^{982}_{32}\mathrm{Ge}$  &  $1.3  \times  10^{14}$  &  $13.8$  &  $8.62$  &  $-0.4$  &  $0.0011$  &  NP
    \\  
    \hline
  \end{tabular}

  \caption{Quantities relevant to the pinning force calculation obtained with the microscopic semi-classical approach in Ref.~\cite{Donati:2004gnw}, where the Argonne potential is used for the nuclear pairing interaction. $\rho$, $R_{\mathrm{WS}}$, and $\xi$ are the mass density, Wigner-Seitz radius, and coherence length, respectively. 
    } 
  \label{tab:vn-force_simulation_Micro-Argonne}
  \end{table}

\begin{table}
  \renewcommand{\arraystretch}{1.5}
  \centering
  \begin{tabular}{ c || c | c | c  c  c | c | c }
    \hline
    zone  &  Element  &  $\rho$ &  $R_{\rm  WS}$  &  $\xi$  &  $E_{\rm  pin}$  &  $f_{\rm  pin}$  &  config
    \\
    &&  [g  cm$^{-3}$]  &  [fm]  &  [fm]  &  [MeV]  &  [MeV  fm$^{-2}$]  &
    \\
    \hline
    \hline
    1  &  $^{320}_{40}\mathrm{Zr}$  &  $1.5  \times  10^{12}$  &  $44.0$  &  $7.76$  &  ---  &  ---  &  IP
    \\
    2  &  $^{1100}_{50}\mathrm{Sn}$  &  $9.6  \times  10^{12}$  &  $35.5$  &  $4.07$  &  ---  &  ---  &  IP
    \\
    3  &  $^{1800}_{50}\mathrm{Sn}$  &  $3.4  \times  10^{13}$  &  $27.0$  &  $3.93$  &  ---  &  ---  &  IP
    \\
    4  &  $^{1500}_{40}\mathrm{Zr}$  &  $7.8  \times  10^{13}$  &  $19.4$  &  $7.78$  &  $-7.5$  &  $0.010$  &  NP
    \\  
    5  &  $^{982}_{32}\mathrm{Ge}$  &  $1.3  \times  10^{14}$  &  $13.8$  &  $8.62$  &  $-5.9$  &  $0.015$  &  NP
    \\  
    \hline
  \end{tabular}
  \caption{Quantities relevant to the pinning force calculation obtained with the microscopic semi-classical approach in Ref.~\cite{Donati:2004gnw}, where the Gogny potential is used for the nuclear pairing interaction. 
    } 
  \label{tab:vn-force_simulation_Micro-Gogny}
  \end{table}

\begin{table}
  \renewcommand{\arraystretch}{1.5}
  \centering
  \begin{tabular}{ c || c | c  c | c  c  c | c | c }
    \hline
    zone  &  Element  &  $\rho$ &  $n$ &  $R_{\rm  WS}$  &  $\xi$  &  $E_{\rm  pin}$  &  $f_{\rm  pin}$  &  config
    \\
    &&  [g  cm$^{-3}$]  &  [fm$^{-3}$]  &  [fm]  &  [fm]  &  [MeV]  &  [MeV  fm$^{-2}$]  &
    \\
    \hline
    \hline
    1a  &  $^{320}_{40}\mathrm{Zr}$  &  $1.7  \times  10^{12}$  &  $0.001$  &  $43.3$  &  $4.43$  &  $-1.08$  &  $0.00029$  &  NP
    \\
    1b  &  $^{320}_{40}\mathrm{Zr}$  &  $3.4  \times  10^{12}$  &  $0.002$  &  $40.0$  &  $4.21$  &  $-1.20$  &  $0.00038$  &  NP
    \\
    1c  &  $^{320}_{40}\mathrm{Zr}$  &  $6.7  \times  10^{12}$  &  $0.004$  &  $36.9$  &  $3.93$  &  ---  &  ---  &  IP
    \\
    2a  &  $^{1100}_{50}\mathrm{Sn}$  &  $1.3  \times  10^{13}$  &  $0.008$  &  $33.0$  &  $4.04$  &  ---  &  ---  &  IP
    \\
    2b  &  $^{1100}_{50}\mathrm{Sn}$  &  $1.8  \times  10^{13}$  &  $0.011$  &  $31.0$  &  $4.12$  &  ---  &  ---  &  IP
    \\
    2c  &  $^{1100}_{50}\mathrm{Sn}$  &  $2.8  \times  10^{13}$  &  $0.017$  &  $28.0$  &  $4.70$  &  ---  &  ---  &  IP
    \\
    3a  &  $^{1800}_{50}\mathrm{Sn}$  &  $4.3  \times  10^{13}$  &  $0.026$  &  $24.5$  &  $6.05$  &  ---  &  ---  &  IP
    \\  
    3b  &  $^{1800}_{50}\mathrm{Sn}$  &  $6.2  \times  10^{13}$  &  $0.037$  &  $21.4$  &  $8.75$  &  $-0.41$  &  $0.00045$  &  NP
    \\  
    \hline
  \end{tabular}
  \caption{
     Relevant quantities for the pinning force calculation obtained with the microscopic quantum approach in Ref.~\cite{Avogadro:2008uy}. 
    } 
  \label{tab:vn-force_simulation_Micro-SLy4}
\end{table}

The results for the microscopic quantum approach in Ref.~\cite{Avogadro:2008uy} are summarized in Table~\ref{tab:vn-force_simulation_Micro-SLy4}, where the SLy4 Skyrme interaction is used for the mean-field interaction. The Wigner-Seitz radius $R_{\rm  WS}$ shown in this table is interpolated from the plot in Ref.~\cite{Negele:1971vb}. For the evaluation of $f_{\mathrm{pin}}$, we again use the formula $f_{\mathrm{pin}} = |E_{\mathrm{pin}}|/(2 R_{\mathrm{WS}}^2)$ and show the values only for the nuclear pinning, just for easy comparison with the semi-classical calculations. 

\begin{table}
  \renewcommand{\arraystretch}{1.5}
  \centering
  \begin{tabular}{c  | c  c  c  c  c }
    \hline
    \multirow{2}{*}{zone}  &  &  &  $f_{\rm  pin}$~[MeV fm$^{-2}$] &  &  
    \\
    \cline{2-6}
    &  $L=100R_{\rm  WS}$  &  $L=500R_{\rm  WS}$  &  $L=1000R_{\rm  WS}$  &  $L=2500R_{\rm  WS}$  &  $L=5000R_{\rm  WS}$
    \\ 
    \hline
    \hline
    1  &  $7.63  \times  10^{-6}$  &  $2.29  \times  10^{-6}$  &  $1.49  \times  10^{-6}$  &  $9.30  \times  10^{-7}$  &  $7.68  \times  10^{-7}$
    \\
    2  &  $2.12  \times  10^{-5}$  &  $6.34  \times  10^{-6}$  &  $4.06  \times  10^{-6}$  &  $2.62  \times  10^{-6}$  &  $2.12  \times  10^{-6}$
    \\
    3  &  $1.43  \times  10^{-4}$  &  $4.49  \times  10^{-5}$  &  $2.74  \times  10^{-5}$  &  $1.54  \times  10^{-5}$  &  $1.14  \times  10^{-5}$
    \\
    4  &  $3.84  \times  10^{-4}$  &  $1.10  \times  10^{-4}$  &  $6.89  \times  10^{-5}$  &  $4.23  \times  10^{-5}$  &  $3.32  \times  10^{-5}$
    \\  
    5  &  $7.85  \times  10^{-5}$  &  $2.71  \times  10^{-5}$  &  $1.89  \times  10^{-5}$  &  $1.36  \times  10^{-5}$  &  $1.12  \times  10^{-5}$
    \\  
    \hline
  \end{tabular}
  \caption{
    The pinning force obtained in the semi-classical mesoscopic approach for different values of $L$ over which the forces exerted on a vortex are integrated~\cite{Seveso:2016}. The zones correspond to those in Table~\ref{tab:vn-force_simulation_Micro-Argonne}. 
    } 
  \label{tab:vn-force_SemiClassical-mesoscopic}
\end{table}

In Table~\ref{tab:vn-force_SemiClassical-mesoscopic}, we show the pinning force obtained in the semi-classical mesoscopic approach for different values of $L$ over which the forces exerted on a vortex are integrated~\cite{Seveso:2016}. The zone numbers in the first column correspond to those in Table~\ref{tab:vn-force_simulation_Micro-Argonne}. We show the calculation in which the reduction of the pairing gap due to the polarization effects in the nuclear matter is not included, corresponding to the choice of the reduction factor $\beta = 1$ introduced in Ref.~\cite{Seveso:2016}. Because of the averaging procedure, we find that a larger $L$ results in a smaller value of $f_{\mathrm{pin}}$.  

\begin{table}
  \renewcommand{\arraystretch}{1.5}
  \centering
  {
  \begin{tabular}{ c | c | c | c | c  c  c}
    \hline
    model
    &  zone
    &  $E_{\rm  pin}$
    &  config
    &  &  $f_{\rm  pin}$~[MeV fm$^{-2}$] &  
    \\
    \cline{4-7}
    &
    &  [MeV]  
    &
    &
    $L=1000R_{\rm  WS}$  &  $L=2500R_{\rm  WS}$  &  $L=5000R_{\rm  WS}$
    \\ 
    \hline
    \hline
    \multirow{8}{*}{SLy4}
    &
    1a  &  $-0.72$  &  NP
    &  $1.39  \times  10^{-6}$  &  $7.38  \times  10^{-7}$  &  $5.40  \times  10^{-7}$
    \\
    &
    1b  &  $-0.91$  &  NP
    &  $1.97  \times  10^{-6}$  &  $1.08  \times  10^{-6}$  &  $8.00  \times  10^{-7}$
    \\
    &
    1c  &  $-0.89$  &  NP
    &  $2.20  \times  10^{-6}$  &  $1.19  \times  10^{-6}$  &  $8.74  \times  10^{-7}$
    \\
    &
    2a  &  $2.73$  &  IP  
    &  $5.61  \times  10^{-6}$  &  $3.72  \times  10^{-6}$  &  $2.95  \times  10^{-6}$
    \\
    &
    2b  &  $3.01$  &  IP  
    &  $7.52  \times  10^{-6}$  &  $5.03  \times  10^{-6}$  &  $4.00  \times  10^{-6}$
    \\
    &
    2c  &  $10.00$  &  IP  
    &  $1.47  \times  10^{-5}$  &  $1.01  \times  10^{-5}$  &  $8.08  \times  10^{-6}$
    \\
    &
    3a  &  $11.78$  &  IP 
    &  $3.25  \times  10^{-5}$  &  $2.31  \times  10^{-5}$  &  $1.88  \times  10^{-5}$
    \\
    &
    3b  &  $9.85$  &  IP 
    &  $8.47  \times  10^{-5}$  &  $6.41  \times  10^{-5}$  &  $5.31  \times  10^{-5}$
  %
  %
    \\ 
    \hline
    \hline
    \multirow{8}{*}{SkM*}
    &
    1a  &  $-0.72$  &  NP
    &  $3.61  \times  10^{-7}$  &  $1.60  \times  10^{-7}$  &  $9.50  \times  10^{-8}$
    \\
    &
    1b  &  $-0.91$  &  NP
    &  $2.20  \times  10^{-7}$  &  $8.72  \times  10^{-8}$  &  $4.08  \times  10^{-8}$
    \\
    &
    1c  &  $-0.89$  &  NP
    &  $2.83  \times  10^{-6}$  &  $1.82  \times  10^{-6}$  &  $1.44  \times  10^{-6}$
    \\
    &
    2a  &  $2.73$  &  IP  
    &  $5.68  \times  10^{-6}$  &  $3.73  \times  10^{-6}$  &  $2.93  \times  10^{-6}$
    \\
    &
    2b  &  $3.01$  &  IP  
    &  $7.58  \times  10^{-6}$  &  $5.01  \times  10^{-6}$  &  $4.00  \times  10^{-6}$
    \\
    &
    2c  &  $10.00$  &  IP  
    &  $1.25  \times  10^{-5}$  &  $8.50  \times  10^{-6}$  &  $6.76  \times  10^{-6}$
    \\
    &
    3a  &  $11.78$  &  IP 
    &  $2.54  \times  10^{-5}$  &  $1.80  \times  10^{-5}$  &  $1.45  \times  10^{-5}$
    \\
    &
    3b  &  $9.85$  &  IP 
    &  $8.00  \times  10^{-5}$  &  $5.93  \times  10^{-5}$  &  $4.79  \times  10^{-5}$
    \\
    \hline
  \end{tabular}
  }
  \caption{
    The pinning force obtained in the quantum mesoscopic calculation where the Skytme interactions, SLy4 and SkM*, are used for the mean-field potential~\cite{Klausner:2023ggr}. The zones correspond to those in Table~\ref{tab:vn-force_simulation_Micro-SLy4}. 
    } 
  \label{tab:vn-force_Quantum-mesoscopic}
\end{table}

Table~\ref{tab:vn-force_Quantum-mesoscopic} shows the pinning force obtained in the quantum mesoscopic calculation where the SLy4 and SkM* Skytme interactions are used for the Hartree-Fock calculation~\cite{Klausner:2023ggr}. The zones correspond to those in Table~\ref{tab:vn-force_simulation_Micro-SLy4}. We again show the results obtained without including the polarization effect, i.e., $\beta = 1$ as in the previous case. 

\begin{figure}
  \centering
    \includegraphics[width=.685\textwidth]{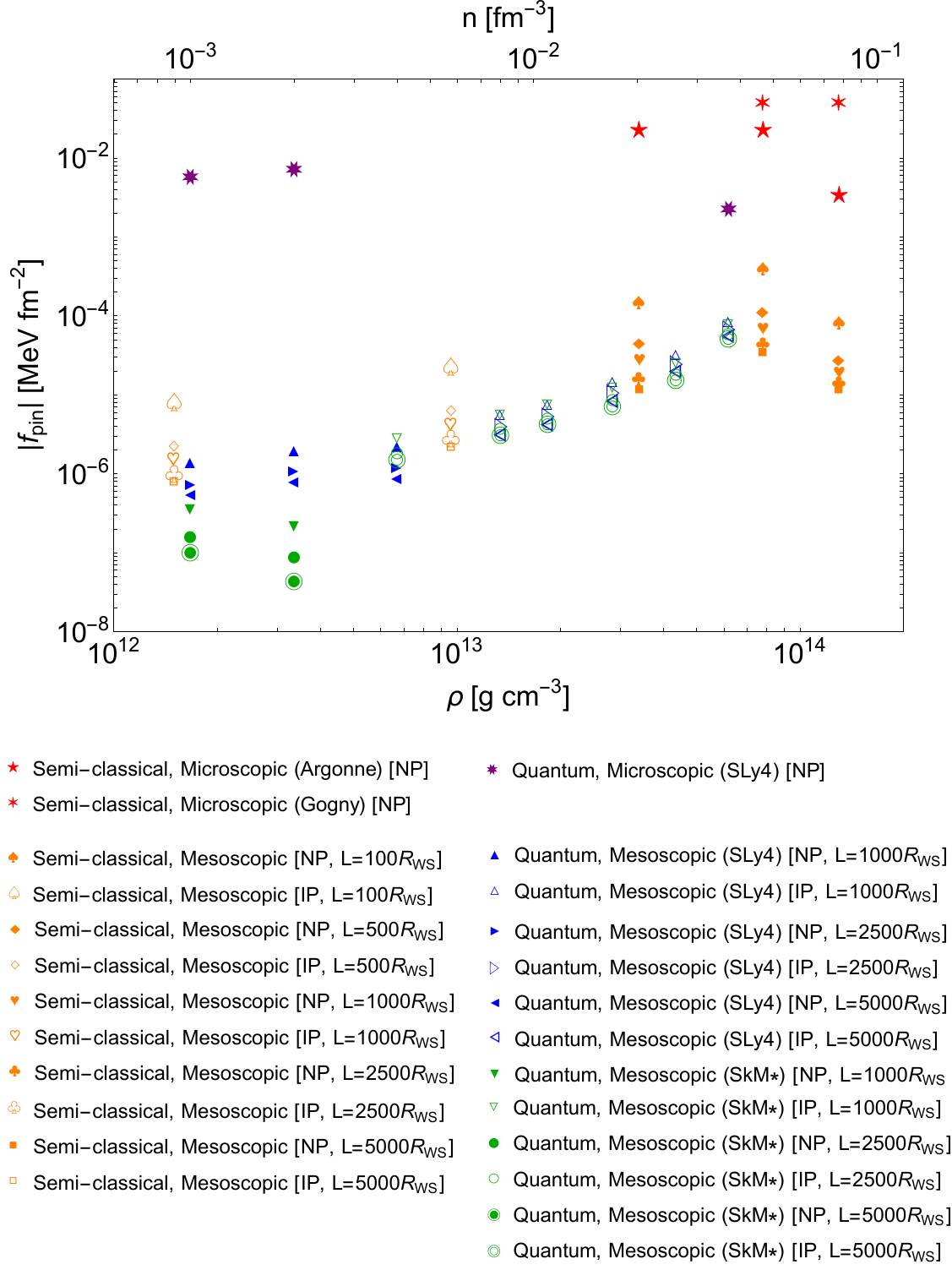}
    \caption{The values of $f_{\rm  pin}$ given in the tables in this appendix against the density $\rho$. The filled and opened markers correspond to the nuclear and interstitial pinnings. 
    }
    \label{fig:rho_vs_fpin}
  \end{figure}

Finally, we plot the values of the pinning force for each density region in Fig.~\ref{fig:rho_vs_fpin}. The filled and opened markers correspond to the nuclear and interstitial pinnings.  As we see, the values of $f_{\rm  pin}$ are distributed in the range $10^{-8}$--$10^{-2}$~MeV $\cdot$ fm$^{-2}$, depending on the evaluation scheme and the selected nuclear potential.

\section{Selection criteria of NS data}
\label{sec:catalog_selection}

We explain how we choose the range of uncertainty of $T_{\rm  s}$ for each NS shown in Table~\ref{tab:plusar_catalogue}.

\begin{itemize}
\item  \textbf{No.~1,~PSR B1706-4}: 
In Ref.~\cite{McGowan:2003sy}, 
the X-ray data of PSR B1706-44 obtained in XMM-Newton is fitted by the BB (blackbody), BB+PL (power law), and atmosphere+PL models, and only BB+PL and atmosphere+PL models result in acceptable $\chi^2$ values. The atmosphere model includes the light-element NS atmosphere (\textit{e.g.} dominated by Hydrogen) and shows large Wien excesses in the high-energy region. Therefore, atmosphere+PL models tend to favor lower temperature and larger radius for the emitting area. We selected the minimum and maximum among the BB+PL and atmosphere+PL models,
$T^\infty  =  (0.48-2.2)  \times  10^6~\mathrm{K}$, 
to include the uncertainty coming from the choice of fitting models.

\item  \textbf{No.~9,~PSR J2043+2740}: Ref.~\cite{Becker:2004gk} studied the XMM-Newton data of PSR J2043+2740. Using the BB + PL model, the upper bound is derived as $T^{\infty}_s  <  6.27  \times  10^5~\mathrm{K}$, where $R_{\rm  NS}=10~\mathrm{km}$ is assumed for the emission radius. 
On the other hand, Ref.~\cite{Zavlin:2004wt} also fitted the X-ray data of the XMM-Newton and obtained an even higher BB temperature, $T^{\infty}_s  \simeq  9  \times  10^5~\mathrm{K}$ with the radiation radius $R^\infty  \simeq  2~\mathrm{km}~\text{\cite{Zavlin:2004wt}}$. Although the fitted radius is smaller than the expected NS radius, it is too large to be interpreted as the magnetic cap radius. Thus, we can not exclude the possibility that this BB temperature corresponds to the emission from the NS surface. To evaluate $T_{\rm  s}$ conservatively, we chose the highest value of the BB temperature as an upper bound. 

\item  \textbf{No.~12,~PSR B0950+08}: In Ref.~\cite{Abramkin:2021fzy}, the optical-UV flux of PSR B0950+08 obtained in the Hubble Space Telescope (HST) far-UV (FUV) detector is analyzed. 
The best-fit temperature is obtained as $T_{\rm  s}  =  (6-12)  \times  10^4~\mathrm{K}$, and we decided to use the proposed value. Note that the conservative upper bound is also derived as $T_{\rm  s}  <  1.7  \times  10^5~\mathrm{K}$ by varying the parameter, such as the ratio of NS radius and distance.

\item  \textbf{No.~17,~PSR J2124-3358}: Ref.~\cite{Rangelov:2016syg} analyzed the optical data from the J2124-3358. 
The BB+PL model gives the following possible range $T_{\rm  s}  \in  [0.5,  2.1]  \times  10^5~\mathrm{K}$ with the uncertainty from the distance. We decided to select the original range, which is almost the same uncertainty added by hand in the way we described in Sec.~\ref{sec:observation}. 

\item  \textbf{No.~19,~RX J2143.0+0654}: In Ref.~\cite{Kaplan:2009au}, the X-ray data in XMM-Newton is fitted using the BB absorption model, and the authors obtain the BB temperature $k_{\rm   B}  T^\infty  =  104.0  \pm  0.4~\mathrm{eV}$ with the BB radius as $R^\infty  =  ( 3.10  \pm  0.04)~\mathrm{km}$, where distance is fixed to be $d=500~\mathrm{pc}$. This value is smaller than the typical NS radius, which implies that this BB temperature is not from the surface but from the small areas around the magnetic caps. In Ref.~\cite{Schwope:2009tv}, the authors fitted the data from Large Binocular Telescope (optical) by combining with the X-ray data in XMM-Newton. Using the BB absorption model, $k_{\rm   B}  T^\infty  =  105.1  \pm  0.9~\mathrm{eV}$ is obtained, 
which is consistent with the result in Ref.~\cite{Kaplan:2009au}. They also perform fitting using the two-component BB model and the hotter (cooler) component is obtained as $k_{\rm   B}  T  =  104~\mathrm{eV}$ ($k_{\rm   B}  T  =  40~\mathrm{eV}$). It is impossible to eliminate uncertainty from the model selection to fit the data from this situation, and thus we choose all the possible ranges of the temperature, $k_{\rm   B}  T_{\rm  s}  =  40$--$106~\mathrm{eV}$. 

\item  \textbf{No.~20,~PSR J0108-1431}: The X-ray data from the direction of J1080-1431 observed in XMM-Newton is fitted by the BB+PL model~\cite{Posselt:2012vy} with $k_{\rm   B}  T  =  110^{+30}_{-10}~\mathrm{eV}$ and $R_{\rm  NS}  =  43^{+16}_{-9}~\mathrm{m}$. This small emission radius implies that this BB component is not the cool surface temperature but the hot magnetic pole component. We can interpret this result as the surface temperature is much cooler than the magnetic pole, and thus, the hot component dominates the observed flux. The latest analysis~\cite{Abramkin:2021tha} analyses both the XMM-Newton and optical data (HST, VLT). In particular, HST F140LP detected thermal emission, and they put the conservative upper bound on the surface temperature as $T_{\rm  s}  <  5.9  \times  10^4~\mathrm{K}$. To derive this conservative bound, they included uncertainty from the parallax distance~\cite{Verbiest:2010tu}. Furthermore, they obtain the value of $T_{\rm  s}$ by assuming the FUV flux is dominated by a thermal component, $T_{\rm  s}  =  27000-55000~\mathrm{K}$. We selected this range to represent the uncertainty. 

\item  \textbf{No.~22,~PSR J2144-3933}: 
The upper bound is obtained for the surface temperature of J2144-3933 using XMM-Newton data (combining with the optical data of Very Large Telescope (VLT))~\cite{Tiengo:2011pw} as $T_{\rm  s}  <  2.3  \times  10^5~\mathrm{K}$. The latest analysis~\cite{Guillot:2019ugf} used deep optical and FUV observation data by HST and derived the upper bound on the surface temperature of J2144-3933. The conservative upper bound on the surface temperature is derived based on the non-detection,
\begin{align}
  T_{\rm  s}  <  4.2  \times  10^4~\mathrm{K},
\end{align}
where a range of NS radius $R_{\rm  NS}  =[11,13]~\mathrm{km}$ and parallax distance $d=172^{+20}_{-15}~\mathrm{pc}$ is considered to estimate the uncertainty~\cite{Verbiest:2010tu}.
In this analysis, NS mass is fixed as $M_{\rm NS}  =  1.4~M_\odot$.

\end{itemize}

Let us also comment on the rejected observational data from our list. 
\begin{itemize}
\renewcommand{\labelitemi}{$-$}
\item  \textbf{PSR B1929+10}: The BB+PL fit is performed for the X-ray data~\cite{Misanovic:2007sb}. However, the magnetic pole component is reported to dominate the temperature because the fitted radiation radius is much smaller than the NS radius. We conclude this data is not appropriate to test the vortex creep heating. 

\item  \textbf{XMMU J1732-344}: We also omit XMMU J1732-344 from our list because the observed value of $|\dot{\Omega}|$ is not determined. Once its pulsation data is fixed, it is worth studying whether the vortex creep heating can explain this data; its thermal emission is expected to exceed the value expected from minimal cooling~\cite{Tian:2008tr,Klochkov:2014ola} with its kinetic time information~\cite{Tian:2008tr}.
\end{itemize}

\bibliographystyle{utphysmod}
\bibliography{references.bib}

\end{document}